\documentclass[reprint,amsmath,aps,prb,twocolumn]{revtex4-1}
\usepackage[utf8]{inputenc}
\usepackage[T1]{fontenc}
\usepackage{graphicx}
\usepackage{dcolumn}
\usepackage{amsmath}
\usepackage{amssymb}
\usepackage{array}
\usepackage{multirow}
\usepackage{calc}
\usepackage{color}
\usepackage{graphicx}
\usepackage{float}
\usepackage{bm}
\usepackage{hyperref}

\usepackage[left,modulo]{lineno}

\newcommand{\icm}{\ensuremath{~\textrm{cm}^{-1}}}
\newcommand{\paa}{(\textit{a,a})}
\newcommand{\pac}{(\textit{a,c})}
\newcommand{\pa}{\textit{a}}
\newcommand{\pc}{\textit{c}}

\newcommand{\degree}{\ensuremath{^\circ}}

\begin{document}

\title{Lattice dynamics of the heavy fermion compound URu$_2$Si$_2$}
\author{J. Buhot}
\author{M.A. M\'easson}
\email{marie_aude.measson[at]univ-paris-diderot.fr}
\author{Y. Gallais}
\author{M. Cazayous}
\author{A. Sacuto}
\affiliation{Laboratoire Mat\'eriaux et Ph\'enom\`enes Quantiques, UMR 7162 CNRS, Universit\'e Paris Diderot, B$\hat{a}$t. Condorcet 75205 Paris Cedex 13, France}
\author{F. Bourdarot}
\author{S. Raymond}
\author{G. Lapertot}
\author{D. Aoki}
\author{L.P. Regnault}
\affiliation{SPSMS, UMR-E CEA / UJF-Grenoble 1, INAC, 38054 Grenoble, France}
\author{A. Ivanov}
\affiliation{Institut Laue Langevin, 38042 Grenoble, France}
\author{P. Piekarz and K. Parlinski}
\affiliation{Institute of Nuclear Physics, Polish Academy of Sciences, 31-342 Krak\`ow, Poland}
\author{D. Legut}
\affiliation{ Nanotechnology Centre, VSB-Technical University of Ostrava, Ostrava, Czech Republic}
\author{C.C. Homes}
\affiliation{Condensed Matter Physics and Materials Science Department, Brookhaven National Laboratory, Upton, New York 11973, USA}
\author{P. Lejay}
\affiliation{Institut N\'eel, CNRS et Universit\'e Joseph Fourier, BP166, F-38042 Grenoble Cedex 9, France}
\author{R.P.S.M. Lobo}
\affiliation{LPEM, PSL Research University, ESPCI-ParisTech, 10 rue Vauquelin, F-75231 Paris Cedex 5, France
CNRS, UMR 8213, F-75005 Paris, France
Sorbonne Universités, UPMC Univ Paris 06, F-75005 Paris, France}

\begin{abstract}

We report a comprehensive investigation of the lattice dynamics of URu$_2$Si$_2$ as a function of temperature using Raman scattering, optical conductivity and inelastic neutron scattering measurements as well as theoretical {\it ab initio} calculations. The main effects on the optical phonon modes are related to Kondo physics. The B$_{1g}$ ($\Gamma_3$ symmetry) phonon mode slightly softens below $\sim$100~K, in connection with the previously reported softening of the elastic constant, $C_{11}-C_{12}$, of the same symmetry, both observations suggesting a B$_{1g}$ symmetry-breaking instability in the Kondo regime. Through optical conductivity, we detect clear signatures of strong electron-phonon coupling, with temperature dependent spectral weight and Fano line shape of some phonon modes. Surprisingly, the line shapes of two phonon modes, E$_u$(1) and A$_{2u}$(2), show opposite temperature dependencies. The A$_{2u}$(2) mode loses its Fano shape below 150 K, whereas the E$_u$(1) mode acquires it below 100~K, in the Kondo cross-over regime. This may point out to momentum-dependent Kondo physics. By inelastic neutron scattering measurements, we have drawn the full dispersion of the phonon modes between 300~K and 2~K. No remarkable temperature dependence has been obtained including through the hidden order transition. {\it Ab initio} calculations with the spin-orbit coupling are in good agreement with the data except for a few low energy branches with propagation in the (a,b) plane.

\end{abstract}

\date{\today}

\maketitle

\section{Introduction}

After almost three decades \cite{palstra_superconducting_1985} of intensive experimental and theoretical research, the nature of the ordered phase found in the Kondo system URu$_2$Si$_2$ at temperature below T$_0$=17.5~K remains to be unraveled \cite{mydosh_colloquium:_2011, mydosh_hidden_2014}. Whereas appearing clearly in the thermodynamic and transport quantities \cite{palstra_superconducting_1985, schlabitz_superconductivity_1986, de_visser_thermal_1986}, the order parameter of this electronic hidden order (HO) state could not be determined by any usual or sophisticated experimental techniques\cite{broholm_magnetic_1991}. Theoretical proposals are numerous, starting from itinerant or localized picture for the 5f electrons \cite{kusunose_hidden_2011-1,haule_arrested_2009,ikeda_emergent_2012,rau_hidden_2012,fujimoto_spin_2011,riseborough_phase_2012,das_imprints_2014,pepin_modulated_2011,chandra_hastatic_2013,elgazzar_hidden_2009}.

Particular features of the HO state have been determined. Inelastic neutron measurements \cite{broholm_magnetic_1991, bourdarot_precise_2010, wiebe_gapped_2007} observe two magnetic excitations with a commensurate wave vector Q$_0$=(1,0,0) and an incommensurate wave vector Q$_1$=(1.4,0,0)$\equiv$(0.6,0,0), the first one being a fingerprint of the HO state \cite{villaume_signature_2008}. A partial Fermi-surface gapping with a strong reduction of the carriers number occurs at T$_0$ \cite{bonn_far-infrared_1988, schoenes_hall-effect_1987}. At higher temperature, a heavy-electron Kondo liquid regime emerges below $\sim$100~K\cite{palstra_superconducting_1985,palstra_magnetic_1986,aynajian_visualizing_2010}. This cross-over temperature, observed in resistivity for instance, has been shown to be drastically reduced under high magnetic field \cite{scheerer_interplay_2012} until the HO state vanishes at $\sim$35~T, suggesting that the Kondo liquid regime is a precursor of the HO state. It is well admitted that a Brillouin zone folding from a body center tetragonal (bct) to a simple tetragonal (st) phase occurs upon entering the HO state \cite{hassinger_similarity_2010,elgazzar_hidden_2009,buhot_a2g_2014}. Recently, various experiments have identified a four-fold symmetry breaking upon entering the HO state  \cite{okazaki_rotational_2011, tonegawa_cyclotron_2012} and orthorhombic static lattice distortion has been reported by Tonegawa et al. \cite{tonegawa_direct_2014}.

The physics of URu$_2$Si$_2$ being mainly electronic, the lattice properties have been hardly investigated. Raman scattering \cite{cooper_raman_1986} mainly reported temperature dependence of the intensity of the fully-symmetric phonon mode and optical conductivity studies were mostly focused on the electronic properties. A more detailed study by ultrasonic measurements versus temperature and under high magnetic field \cite{yanagisawa__2013, kuwahara_lattice_1997} reported a softening of the elastic constant $C_{11}-C_{12}$ below $\sim$ 120~K suggesting a B$_{1g}$-type (or $\Gamma_3$) lattice instability in connection with the Kondo cross-over. Quite recently, an anomalous phonon softening below $T_{0}$ in the [1,1,0] direction has been reported by inelastic neutron scattering \cite{butch_soft_2012}, calling on for further detailed studies.

We report here a comprehensive study of the lattice dynamics of URu$_2$Si$_2$ from 300~K to 2~K; the optical phonon modes have been investigated by Raman scattering (section~\ref{raman}) and infrared (IR) (section~\ref{opt}) spectroscopies, the dispersion of the phonon branches by inelastic neutron scattering (including polarization techniques) (section~\ref{INS}), and {\it ab initio} calculations (section~\ref{sec:theory}) were used for comparison with all measurements.

\section{Phonon modes in URu$_2$Si$_2$ and selection rules}

\begin{table*}[t]
\caption{Raman and Infrared active modes for the Point Group D$_{4h}$ (I4/mmm) with corresponding irreducible representations, symmetry, atom displacements in URu$_2$Si$_2$.}

\renewcommand{\arraystretch}{1.8}
\begin{tabular}{|c|c|c|c|c|}
\hline
Modes     & Irreductible representation & Corresponding symmetry         &  Phonon         & Atom displacements  \\
activity  & in Mulliken's notation      & in Bethe's notation (function) &  Multiplicity   &(direction)          \\
\hline
\multirow{5}{*}{Raman active}& A$_{1g}$   & $\Gamma_{1}^{+}\; (x^2+y^2, z^2)$               & 1 & Si (c-axis)        \\
                             & A$_{2g}$   & $\Gamma_{2}^{+}\; (J_{z}, ixy(x^2+y^2))$                      & 0 & No active phonon   \\
                             & B$_{1g}$   & $\Gamma_{3}^{+}\; (x^2-y^2)$                    & 1 & Ru (c-axis)        \\
                             & B$_{2g}$   & $\Gamma_{4}^{+}\; (xy)$                         & 0 & No active phonon   \\
                             & E$_{g}$    & $\Gamma_{5}^{+}\; (xz,yz),(J_{x},J_{y})$        & 2 & Si+Ru (ab-plane)   \\
\hline
\multirow{2}{*}{IR active}   & A$_{2u}$   & $\Gamma_{2}^{-}\; (z)$                          & 2 & U+Ru+Si (c-axis)   \\
                             & E$_{u}$    & $\Gamma_{5}^{-}\; (x,y)$                        & 2 & U+Ru+Si (ab-plane) \\
\hline
\end{tabular}
\label{tab1}
\end{table*}

\begin{figure}[h]
\centering
\includegraphics[width=1\linewidth]{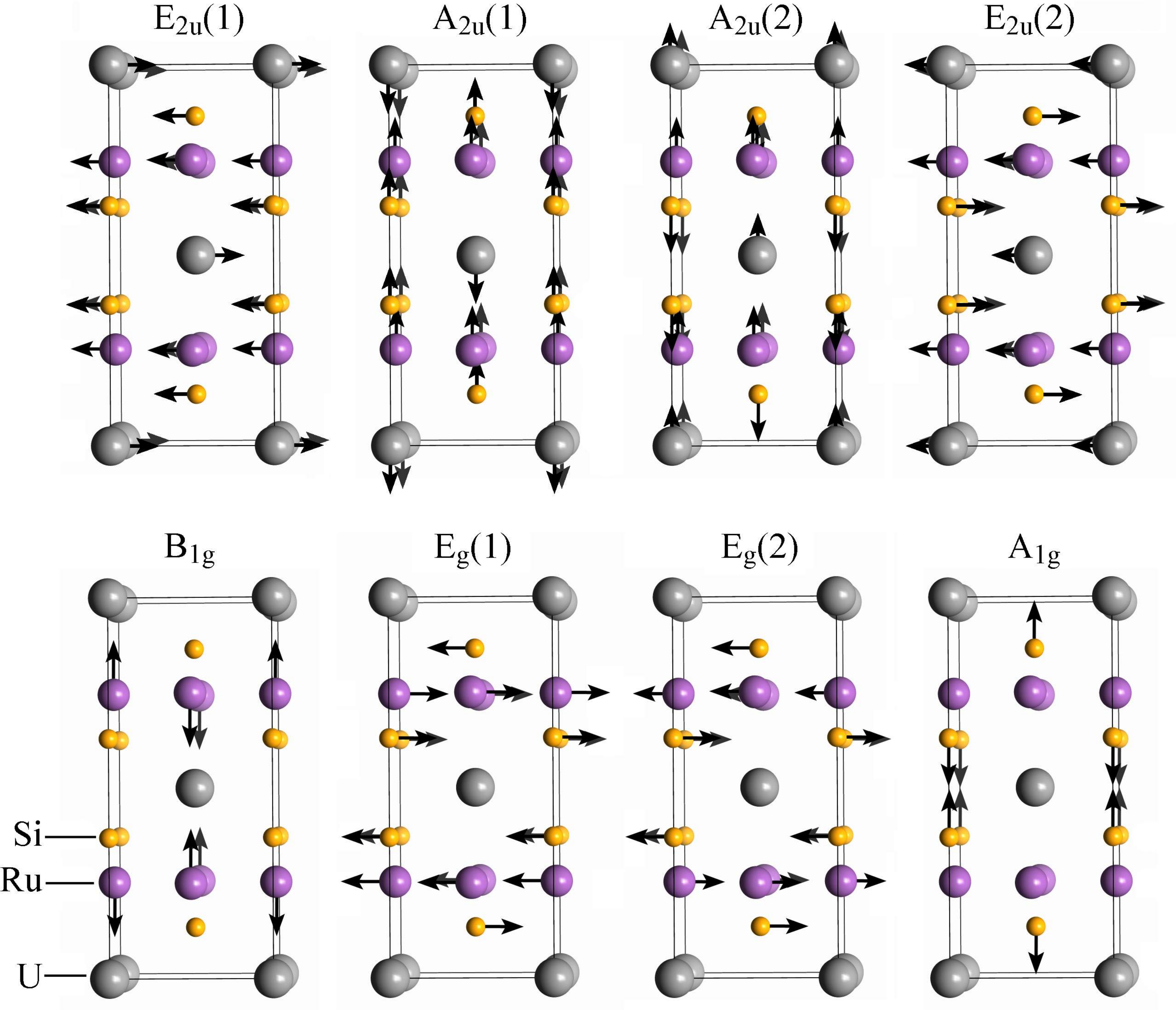}
\caption{(Color online) Atomic displacements sketches of the IR-active optical modes (up) and of the Raman-active optical modes (bottom) of URu$_2$Si$_2$. From the left to the right, the modes have increasing energy.}
\label{fig1}
\end{figure}

The URu$_2$Si$_2$ compound belongs to the tetragonal space group I4/mmm (D$_{4h}$), with the U, Ru and Si atoms located at the 2a, 4d and 4e Wyckoff positions, respectively. From group symmetry analysis\cite{hayes_scattering_2004}, 8 zone center optical phonons are expected, A$_{1g}$, B$_{1g}$, 2E$_{g}$, 2A$_{2u}$ and 2E$_{u}$. Table~\ref{tab1} reports all these phonon modes, their multiplicity and the motion of the atoms involved. Due to inversion center in URu$_2$Si$_2$ elementary cell, the gerade (g) mode are Raman active and the ungerade (u) ones are IR actives. The corresponding atomic displacement patterns are sketched in Figure~\ref{fig1}. A$_{1g}$ and B$_{1g}$ modes involve motions of the Si and Ru atoms, respectively, along the c-axis. The E$_{g}$ modes correspond to the motions of Si and Ru atoms in the ab-plane. The IR active modes involve motions of all atoms along the c-axis for the A$_{2u}$ modes and in the ab-plane for the E$_{u}$ modes.

In optical measurements, all the symmetries have been probed by combining different orientations of the samples and/or different incident ($\vec{e}_i$) and scattered ($\vec{e}_s$) light polarizations. Both Raman and IR scattering probes excitation with a transfered wave vector $\vec{Q}$ close to zero. To obtain the full dispersion of the optic and acoustic phonon branches we have carried out inelastic neutron scattering (INS). Here the transverse or longitudinal character of the phonon modes has been obtained by using different configurations of scattering vector $\vec Q$ as well as by comparison with the theoretical prediction for each branch.

\section{Raman spectroscopy of the phonons}
\label{raman}

\subsection{Methods}

\par
Polarized Raman scattering has been performed in quasi-backscattering geometry with a incident laser line at 532~nm from a solid state laser. We have used a closed-cycle $^{4}$He cryostat with sample in high vacuum (10$^{-6}$~mbar) for the measurements from 8~K to 300~K and a $^{4}$He pumped cryostat with the sample in exchange gas for measurements below 8~K or under magnetic field up to 10~T. By comparing Stokes and anti-Stokes Raman spectra and via the evolution of phonon frequencies with incident laser power, we have estimated the laser heating of the samples at +1.3~K/mW and +1~K/mW for the samples in high vacuum and in exchange gas, respectively. Typical laser power of 5~mW was used. The scattered light was analyzed by a Jobin Yvon T64000 triple substractive grating spectrometer equipped with a cooled CCD detector. In the triple substractive configuration we used, the resolution of the spectrometer is 2.5~cm$^{-1}$. For large energy scale measurements (up to 3000~cm$^{-1}$), the spectrometer was used in the simple grating configuration, with a lower resolution. The contribution of the Bose factor has been removed for all spectra.

\subsection{Sample preparation}

\par
\begin{figure}[h]
\centering
\includegraphics[width=1\linewidth]{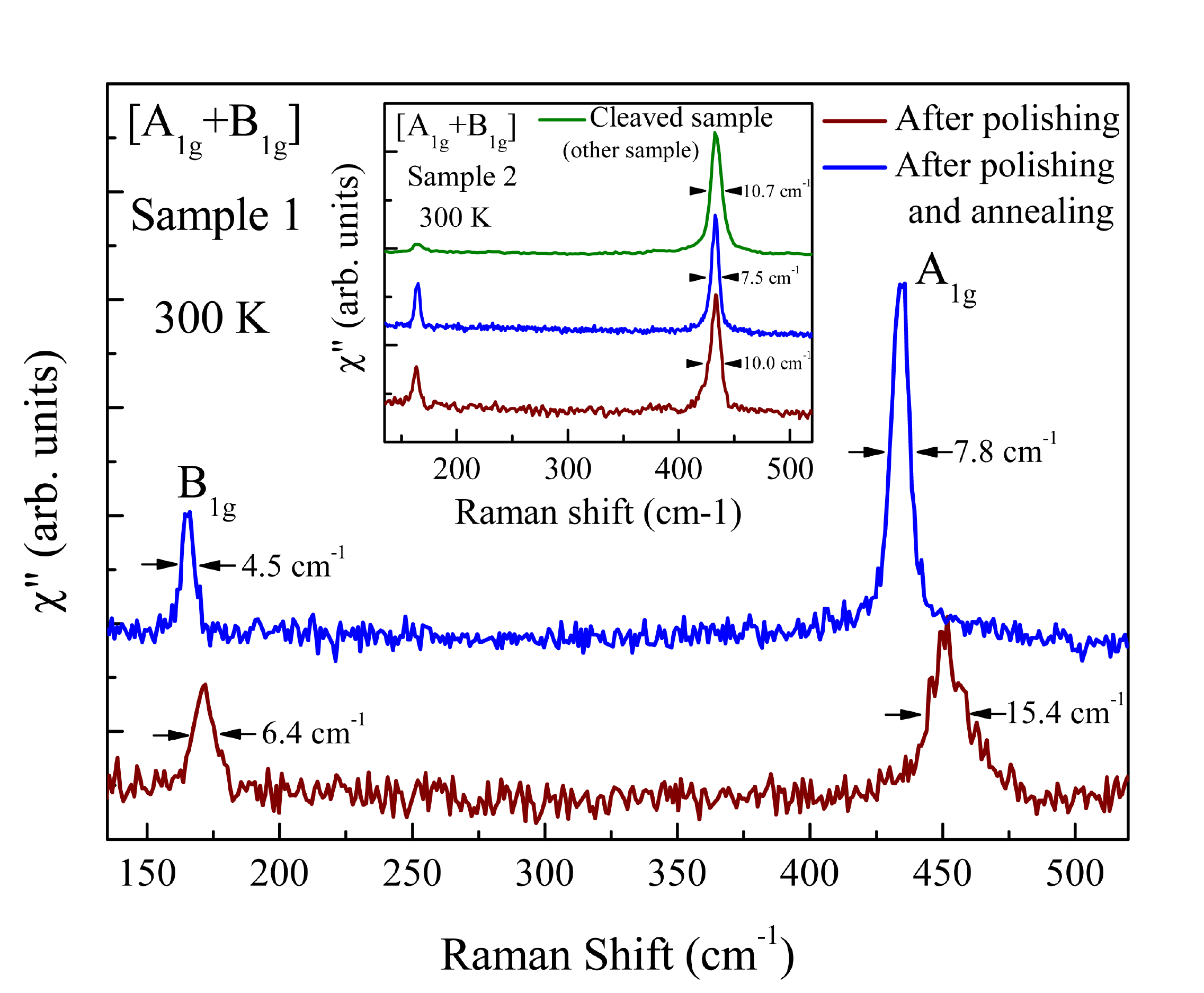}
\caption{(Color online) Raman spectra of URu$_2$Si$_2$ for sample 1 after polishing along the (a,c) plane and annealing. Inset: Raman spectra for sample 2 after polishing along the (a,a) plane, annealing and for another sample cleaved along the (a,a) plane.}
\label{fig0}
\end{figure}

The URu$_2$Si$_2$ single crystals were grown by the Czochralski method using a tetra-arc furnace \cite{aoki_field_2010}. Two samples, from the same batch, were prepared. The initial residual resistivity ratio of the samples is about 50. Samples 1 and 2 were polished along the (a,c) and (a,a) planes, respectively. By cleaving, we obtained samples along the (a,a) plane. The E$_{g}$ phonon modes can be probed only in (a,c) plane ($\vec E$//(a,c)), so only on polished sample. Figure~\ref{fig0} shows the A$_{1g}$ and B$_{1g}$ phonon modes after polishing then after annealing, for sample 1 and for sample 2 (see inset). After polishing along the (a,c) plane (sample 1), both phonon modes are shifted by about 4\% to higher energy and broadened. Most probably, the stress induced by polishing gives rise to such hardening and to the shortening of their lifetime. However, no such stress effect have been observed on the sample polishing along (a,a) plane (see inset of Figure~\ref{fig0}). In order to release the stress induced by polishing, both samples have been annealed for two days at 950~\degree under ultra high vacuum. This process has shifted down the phonon modes and it has clearly sharpened them. The final position and width are similar to what is measured on the cleaved sample of the same batch\cite{buhot_raman_2013} (see inset of Fig.\ref{fig0}) (Note that even sharper A$_{1g}$ phonon mode with a full width at half maximum (FWHM) at 4~K of 4\icm have been measured on cleaved samples from another batch). The E$_g$ modes measured on final sample 1 are very sharp, comparable to the resolution of the spectrometer at low temperature.

\subsection{Anharmonic model}
\label{raman:model}

The temperature dependence of the FWHM, $\Gamma_{ph}$, and energy, $\omega_{ph}$, of the phonon mode is usually described by a simple symmetric anharmonic decay model, i.e. decay of an optical phonon into two acoustic modes with identical frequencies and opposite momenta\cite{klemens_anharmonic_1966, menendez_temperature_1984} :

\begin{equation}
\omega_{ph}(T)=\omega_{0}-C\coth(\frac{\hbar\omega_{0}}{4k_{B}T})	
\end{equation}
\begin{equation}
\Gamma_{ph}(T)=\Gamma_{0}-\Gamma\coth(\frac{\hbar\omega_{0}}{4k_{B}T})	
\end{equation}

where $C$ and $\Gamma$ are positive constants, $\omega_{0}$ is the bare phonon frequency, and $\Gamma_{0}$ a residual (temperature independent) linewidth originating from sample imperfections.
With this purely phononic effect, upon cooling down, the energy and the width of the phonon mode hardens and decreases, respectively, before saturating.

In addition, electron-phonon coupling can also induce renormalization of the frequency and width as well as a change of the shape of the phonon mode. In a simple model~\cite{fano_effects_1961, cardona_light_1982}, the width change is directly related to the electronic density at the energy of the phonon mode and the frequency relies on the full electronic spectrum. Generally, a loss of electronic density of states produces a narrowing of the phonon mode. Such coupling can also induce an asymmetric Fano shape of the phonon. If this last effect remains negligible, the line shape of the phonon can be described by a Lorentzian profile, i.e. : $$\chi''(\omega) \propto \frac{\Gamma_{ph}/2}{(\omega-\omega_{ph})^2-(\Gamma_{ph}/2)^2}$$

\subsection{Results}
\label{raman:results}
\par
\begin{figure}[h]
\centering
\includegraphics[width=1\linewidth]{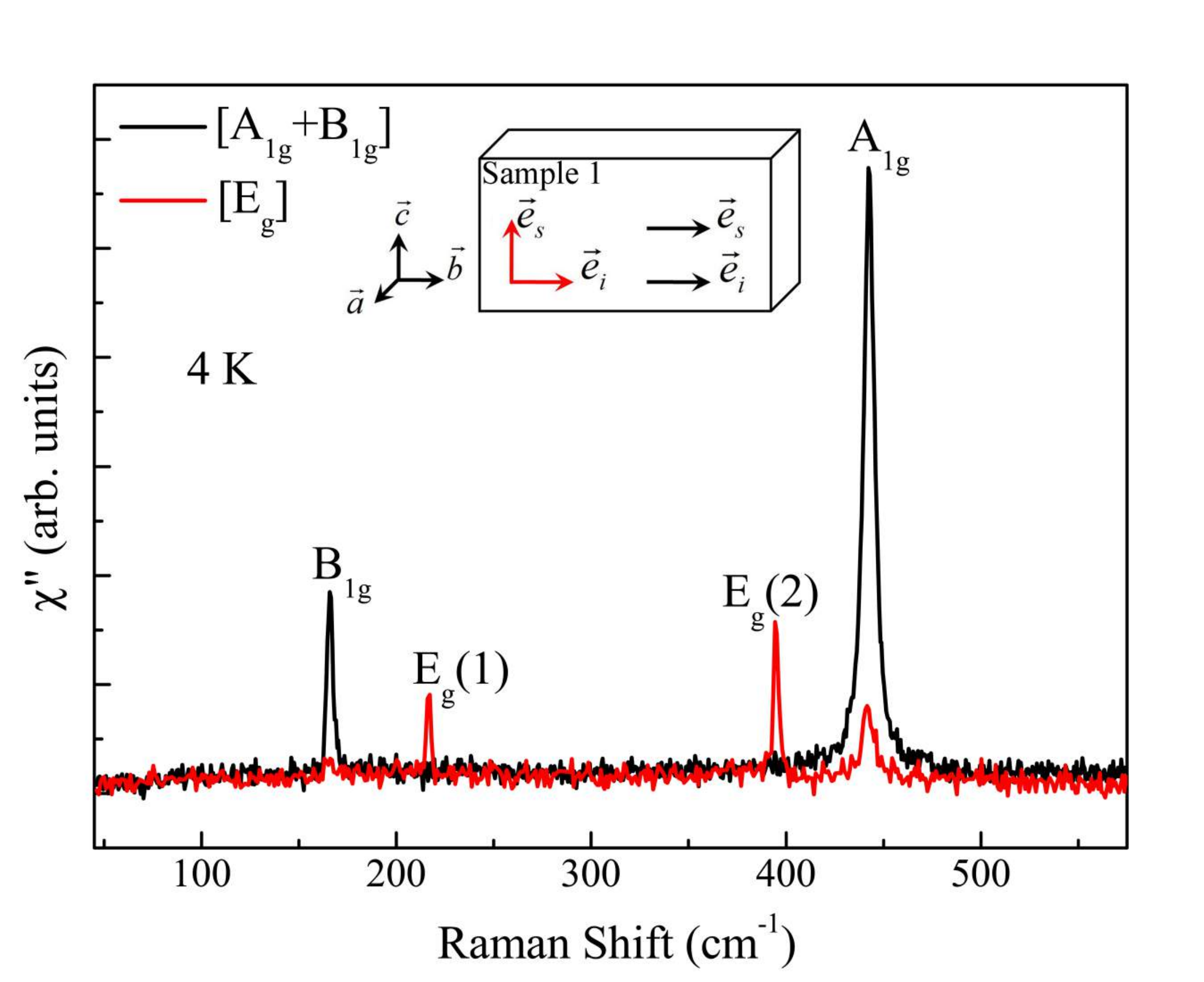}
\caption{(Color online) Typical Raman spectra measured at 4~K in cross and parallel polarization of the light and for sample 1. A scheme of sample 1 with the orientation of the polarizations of the incident $\vec{e}_{i}$ and scattering $\vec{e}_{s}$ light is shown.}
\label{fig2}
\end{figure}

Figure~\ref{fig2} shows typical Raman spectra obtained at 4~K on sample 1 after polishing and annealing. A$_{1g}$ and B$_{1g}$ phonon modes are visible in parallel polarization and E$_g$ modes in cross-polarization. We observe a leakage of the A$_{1g}$ mode in cross-polarization due to a weak crystal misalignment. At 300~K, the two E$_g$ modes are seen at 213 and 391 cm$^{-1}$, and A$_{1g}$ and B$_{1g}$ at respectively, 434 and 163 cm$^{-1}$. The B$_{1g}$ and both E$_g$ phonon modes are sharper (with a FWHM of 3.2 cm$^{-1}$, 2.8 cm$^{-1}$ and 1.8 cm$^{-1}$ at 4K, respectively) than the A$_{1g}$ phonon mode (FWHM=6.6 cm$^{-1}$ at 4K). All phonons have a Lorentzian line shape.

\begin{figure}[h]
\includegraphics[width=1\linewidth]{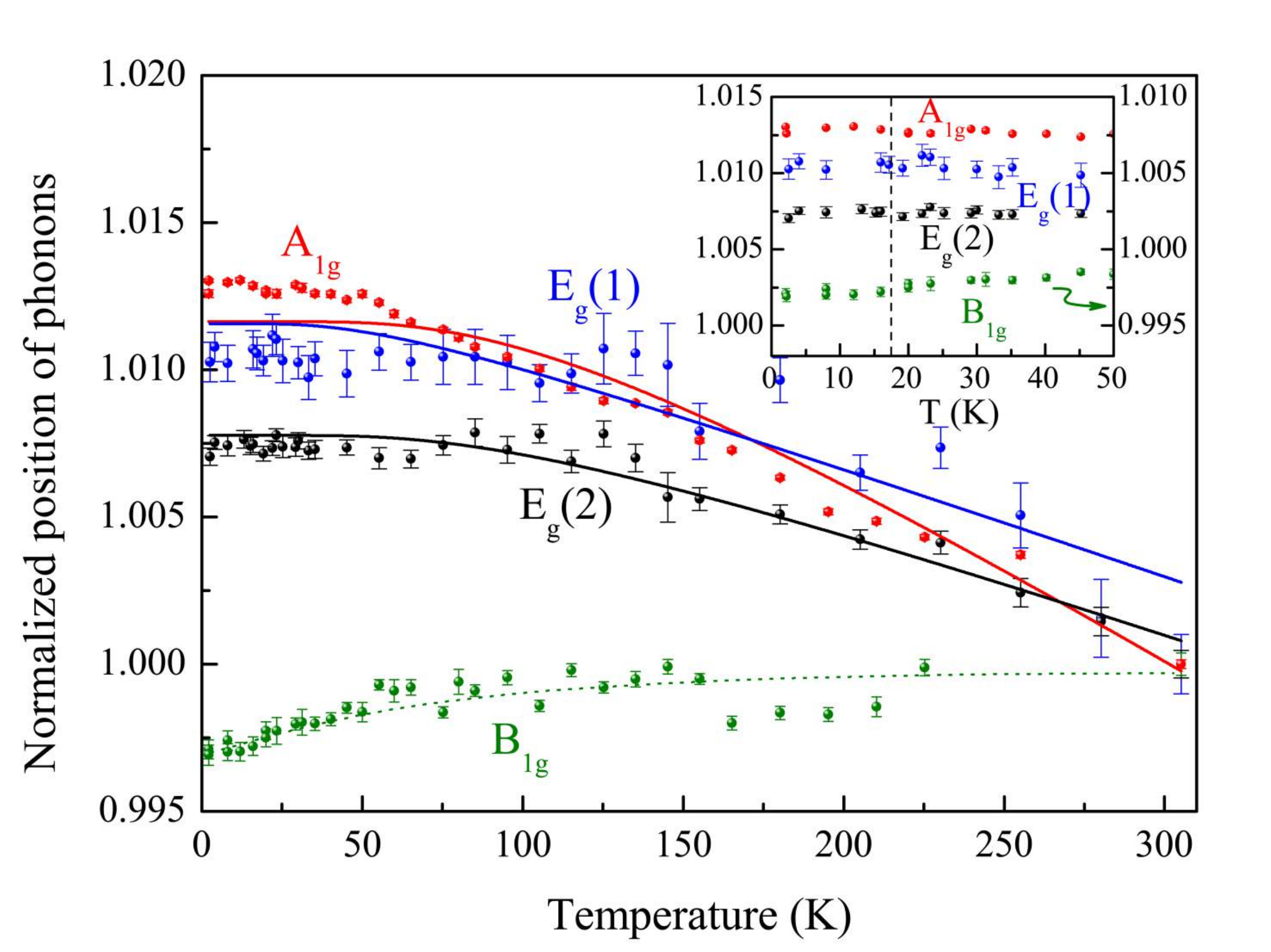}
\caption{(Color online) Temperature dependence of the position of the A$_{1g}$, E$_g$(1), E$_g$(2) and B$_{1g}$ phonon modes normalized to their value at 300~K. A$_{1g}$ and E$_{g}$ phonon modes have a typical temperature dependence whereas the B$_{1g}$ phonon exhibits an unusual softening when cooling down below 100~K. Error bars have been extracted from the Lorentzian fits. Lines are fits with a single anharmonic model (see section~\ref{raman:model}). The green dot line is a guide to the eyes. Inset : zoom below 50~K across the hidden order transition marked by the vertical black dashed line. No particular anomaly is measured at T$_{0}$.}
\label{fig3}
\end{figure}

\begin{figure}[h]
\centering
\includegraphics[width=1\linewidth]{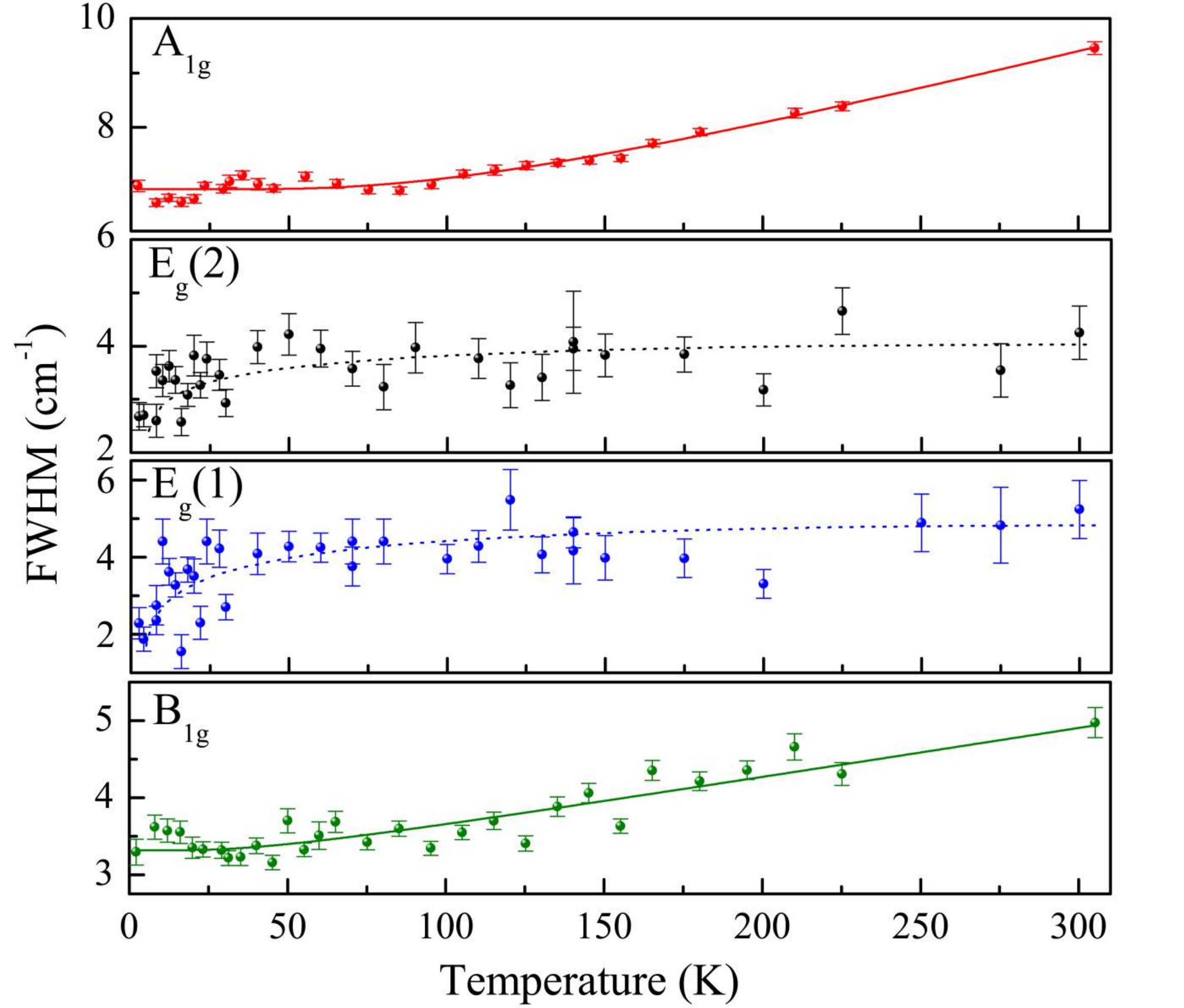}
\caption{(Color online) Temperature dependence of the FWHM of A$_{1g}$, E$_g$(1), E$_g$(2) and B$_{1g}$ phonon modes. Lines are fits with a single anharmonic model (see section~\ref{raman:model}). Dot lines are guides to the eyes.}
\label{fig4}
\end{figure}
\par

Figure~\ref{fig3} and \ref{fig4} present the temperature dependence of the energies and FWHM of the Raman-active phonon modes. The energies are normalized to their value at 300~K.
We have investigated precisely the energy of the phonons through the hidden order transition but no particular effect has been observed within our accuracy (see inset of Figure~\ref{fig3}).
The energy of the A$_{1g}$ mode increases with decreasing temperature before saturating at 1.25$\%$ higher energy than at 300~K. It narrows upon cooling before saturating. The general temperature dependence of this mode is naturally explained by anharmonic effects (full line are fits with the anharmonic model described in section~\ref{raman:model}). Other ingredients, like anharmonic effect of higher rank (four-phonon process), would be necessary to accurately fit the data.

The temperature dependence of the A$_{1g}$ mode energy is consistent with previous Raman experiments~\cite{cooper_magnetic_1987}. No particular change of the integrated intensity of the A$_{1g}$ phonon mode has been detected contrary to what Cooper et al. have reported\cite{cooper_magnetic_1987}. Nor do we see any abrupt increase of the linewidth of A$_{1g}$ below 20~K contrary to what is reported by Lampakis et al.~\cite{lampakis_raman_2006}.

Whereas both E$_{g}$ modes exhibit usual increasing energy when cooling down, their FWHM are almost constant in all the temperature range with a slight sharpening below $\sim20-30$~K. Within our accuracy, the increase of the lifetime of these phonons might be concomitant with the electronic gap opening at T$_0$ observed by optical conductivity \cite{bonn_far-infrared_1988, nagel_optical_2012,guo_hybridization_2012,levallois_hybridization_2011} and Raman scattering \cite{buhot_a2g_2014}. If so, a simple electron-phonon coupling model~\cite{fano_effects_1961, cardona_light_1982} would qualitatively explain such behavior.

\par
Intriguingly, whereas the FWHM of the B$_{1g}$ phonon shows the usual temperature evolution with narrowing when temperature decreases, its energy remains constant down to about 100~K before softening by about 0.5$\%$ below. This softening occurs in the temperature range of the Kondo cross-over and upon entering the Kondo liquid regime below the Kondo temperature reported between 70~K\cite{palstra_superconducting_1985,palstra_magnetic_1986} and 120~K \cite{aynajian_visualizing_2010}. Clearly, the temperature dependence of the energy of the B$_{1g}$ phonon cannot be reproduced by the simple anharmonic model. Moreover, contrary to what we observe, a simple electron-phonon coupling model generally predicts a change of the width as an initial effect on top of which the energy of phonon can be modified. We will further discuss this B$_{1g}$-symmetry softening in section~\ref{raman:disc}.

\par
\begin{figure}[!h]
\centering
\includegraphics[width=1\linewidth]{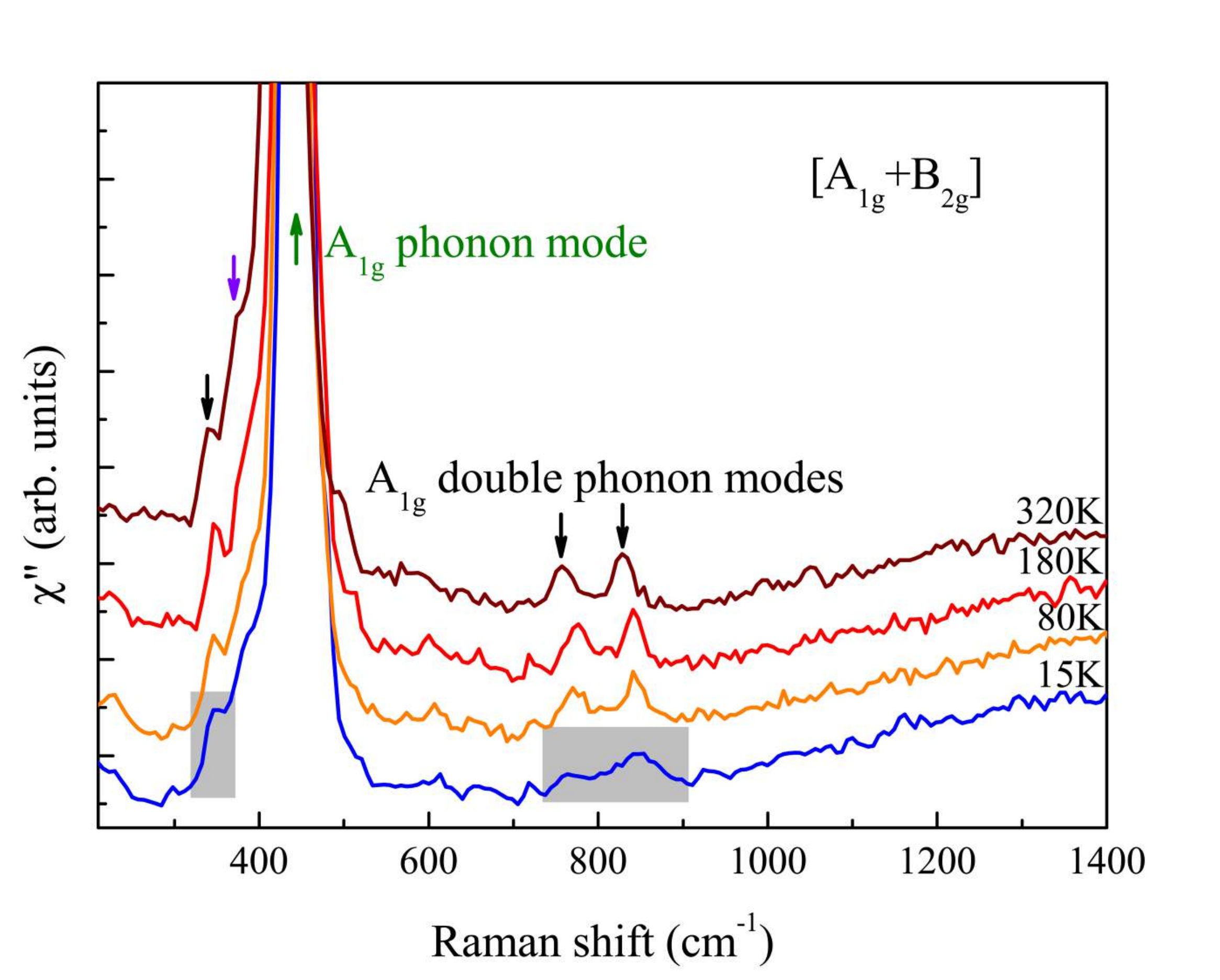}
\caption{(Color online) Large energy scale Raman spectra in A$_{1g}$ + B$_{2g}$ symmetry at various temperatures. Spectra have been shifted for clarity. Green arrow points to the center of the A$_{1g}$ phonon mode. A leakage of the E$_{g}$(2) phonon mode is measured at about 394 cm$^{-1}$ (purple arrow). Three new broad peaks at about 350cm$^{-1}$ and 800 cm$^{-1}$ are indicated by black arrows. Grey areas indicate the energy range for double phonon processes scattering according to theoretical calculations (see section~\ref{sec:theory} and Fig.\ref{fig9}). The three double phonon peaks are of pure A$_{1g}$ symmetry.}
\label{fig5}
\end{figure}

 A large energy scale investigation in all symmetries reveals three new peaks in A$_{1g}$ symmetry, as shown Figure~\ref{fig5}. In A$_{1g}$ + B$_{2g}$ symmetry three broad peaks appear at 350, 760 and 832~cm$^{-1}$ with a linewidth of 21, 37 and 37~cm$^{-1}$, respectively. A slight hardening is observed with decreasing temperature. These new excitations are of pure A$_{1g}$ symmetry as they are not observed in the B$_{1g}$ + B$_{2g}$ symmetry (not shown). These broad features could be due to crystal electric field excitations or double phonon processes. Because of a good agreement with theoretical calculation (see section~\ref{sec:theory}), we attributed them to this last process. Indeed as shown in Figure~\ref{fig5} and \ref{fig9}, the energy ranges of the peaks (grey areas) are consistent with double excitations, either pure or mixed, of the B$_{1g}$, E$_g$ or A$_{1g}$ branches.
\par
Finally, we have probed all phononic excitations under high magnetic field up to 10~T (not shown). No effect has been observed.

\subsection{Discussion}
\label{raman:disc}

Our most noticeable result of the Raman scattering investigation is the unusual temperature dependence of the energy of the B$_{1g}$ phonon mode with softening below the Kondo temperature.
The B$_{1g}$ mode breaks the four-fold rotational symmetry. So the behavior of this B$_{1g}$ phonon suggests a tendency toward lattice instability with orthorhombic distortion. Four-fold symmetry breaking and even orthorhombic distortion upon entering the HO state have been reported by various experiments \cite{okazaki_rotational_2011, tonegawa_cyclotron_2012, tonegawa_direct_2014}. However, the symmetry broken here is B$_{2g}$, i.e. 45\degree~ from the B$_{1g}$ symmetry, both being in the (a,b) plane. Moreover, the temperature ranges are claimed to be different. Clearly, there is no direct relationship between these measurements and our Raman scattering result. The tendency toward lattice instability with B$_{1g}$ symmetry is most probably related with the Kondo physics. Interestingly, similar softening effect (of 0.7\%) of the elastic constant $(C_{11}-C_{12})/2$ in the same symmetry (B$_{1g}\equiv\Gamma_3$) has been reported below 120~K by ultrasound velocity measurements\cite{yanagisawa__2013}. In addition, they show that this effect disappears when high magnetic field of 35~T is applied along the c axis. At this magnetic field, the coherence temperatures are strongly reduced concomitantly with the vanishing of the hidden order phase\cite{scheerer_interplay_2012}. The softening of the elastic constant $(C_{11}-C_{12})/2$ has been related to the emergence of the hybridized electronic state between the 5f electron and the conduction electrons (s or d) and associated to a symmetry-breaking band instability. Both results, on acoustic (ultrasound experiment) and optical phonon (Raman experiment) modes, are nicely consistent and point to a B$_{1g}$ symmetry-breaking instability upon entering into the Kondo regime of URu$_2$Si$_2$. While the acoustic phonon modes involve the motion of all atoms, the B$_{1g}$ mode involves only the Ru atoms. This may suggest that the electronic environment of the Ru atoms are particularly affected by the Kondo physics.

On the basis of our inelastic neutron scattering study and theoretical calculations, we conclude that two origins for this small B$_{1g}$ lattice instability are unlikely. First, our theoretical calculations including global anharmonic effects from purely phononic origin (See Section \ref{sec:theory} and Figure~\ref{fig10}) indicates that none of the phonon branches, except the A$_{2u}$ one, are strongly affected by these anharmonic effects. Secondly, by following the full dispersion of the "B$_{1g}$" branch as well as magnetic excitations by inelastic neutron scattering (See \ref{INS}) we show that the k-dependence of the phonon is smooth going through the minima in the magnetic dispersion ($\vec{Q_0}$ and $\vec{Q_1}$). This does not give any indication of magneto-elastic coupling. Finally, as the B$_{1g}$ mode is not affected by the large loss of carriers upon entering the HO state, as the phenomenon which induces the unusual B$_{1g}$ energy behavior does not involve any noticeable change of its FWHM, a complex electron-phonon coupling related to the Kondo physics is certainly in play for this B$_{1g}$ mode.

\section{Optical conductivity of phonons}
\label{opt}
\subsection{Methods}

Unpolarized optical reflectivity was measured on a cleaved
$2 \times 3 \text{mm}^2$ \paa\ plane. The \pc-axis
reflectivity was taken with appropriate polarizers on sample 1 (see section~\ref{raman}), an optically
polished $1 \times 1.5 \text{mm}^2$ \pac\ surface. We checked
that the data along the \pa\ direction at 5 K was identical
for unpolarized measurements on the \paa\ plane and
\pa-polarized measurements on the \pac\ surface.
Spectra were recorded at several temperatures from 5 K to 300 K,
between 20\icm\ (2.5 meV) and $12\,000\icm$ (1.5 eV). This data was extended
to $40\,000\icm$ (5 eV) at room
temperature. To obtain the absolute reflectivity we utilized an
\textit{in-situ} overfilling (gold or aluminum) evaporation technique.\cite{homes_technique_1993}
In order to obtain the optical conductivity from Kramers-Kronig,
we took a Hagen-Rubens extrapolation below our lowest measured frequency.
Above $40\,000\icm$, we utilized the data by \citeauthor{degiorgi_electrodynamic_1997}\cite{degiorgi_electrodynamic_1997}
up to $100\,000\icm$ (12 eV), followed by a $\omega^{-4}$ free electron termination.

%
%

\subsection{Data modeling}

In this paper, we are interested in the real part of the optical
conductivity ($\sigma_1$), which is built upon the sum of the
contributions from individual excitations:
\begin{equation}
\sigma_1 = \sigma_1^c + \sum_b \sigma_1^b + \sum_p \sigma_1^p \, ,
\label{sigma}
\end{equation}
where the superscripts $c$, $b$ and $p$ stand for contributions from
coherent mobile charge carriers, interband transitions, and phonons,
respectively.

We describe the coherent part by a Drude conductivity in the form
\begin{equation}
\sigma_1^c(\omega)=\frac{2\pi}{Z_0} \frac{\Omega^{2}_{p}}{\omega^{2}\tau+\frac{1}{\tau}} \, ,
\label{Drude}
\end{equation}
where $\Omega_{p}$ is a plasma frequency, $\tau^{-1}$ the scattering rate,
and $Z_0$ the vacuum impedance. Both, interband and phonon contributions
can be modeled by Lorentz oscillators
\begin{equation}
\sigma_1^{b,p}(\omega)=\frac{2\pi}{Z_{0}} \frac{\gamma \, \omega^2 \, \Delta \varepsilon \, \Omega^2}{(\Omega^2-\omega^2)^2+\gamma^2 \omega^2} \, ,
\label{Lorentz}
\end{equation}
where each contribution has a resonance frequency $\Omega$, a linewidth $\gamma$, and a dielectric oscillator strength $\Delta \varepsilon$.

The Lorentz model does not take into account coupling between localized
(\textit{e.g.} phonon) states and the continuum. \citeauthor{fano_effects_1961}\cite{fano_effects_1961}
described phonon lineshapes in a conducting medium, when an electron-phonon coupling
exists. Here we adopt the
formalism proposed by \citeauthor{davis_interaction_1977}\cite{davis_interaction_1977}, who
generalized Fano's approach to multiple discrete states:
\begin{equation}
\sigma_1^F = \frac{2 \pi}{Z_0} R \omega
\left\{
\frac{\left[q \gamma + (\omega - \Omega)\right]^2}
      {\gamma^2 + (\omega - \Omega)^2} - 1
\right\} \, ,
\label{Fano}
\end{equation}
where two new quantities are introduced. $R$ is a renormalization
parameter that takes into account the transition rate between continuum
and localized states. But more interesting is the Fano-Breit-Wigner $q^{-2}$, which
vanishes when the electron-phonon interaction disappears. It
probes the continuum density of states at the phonon frequency.

%
%

\subsection{Results \& Discussion}

\begin{figure*}
\includegraphics[width=2\columnwidth]{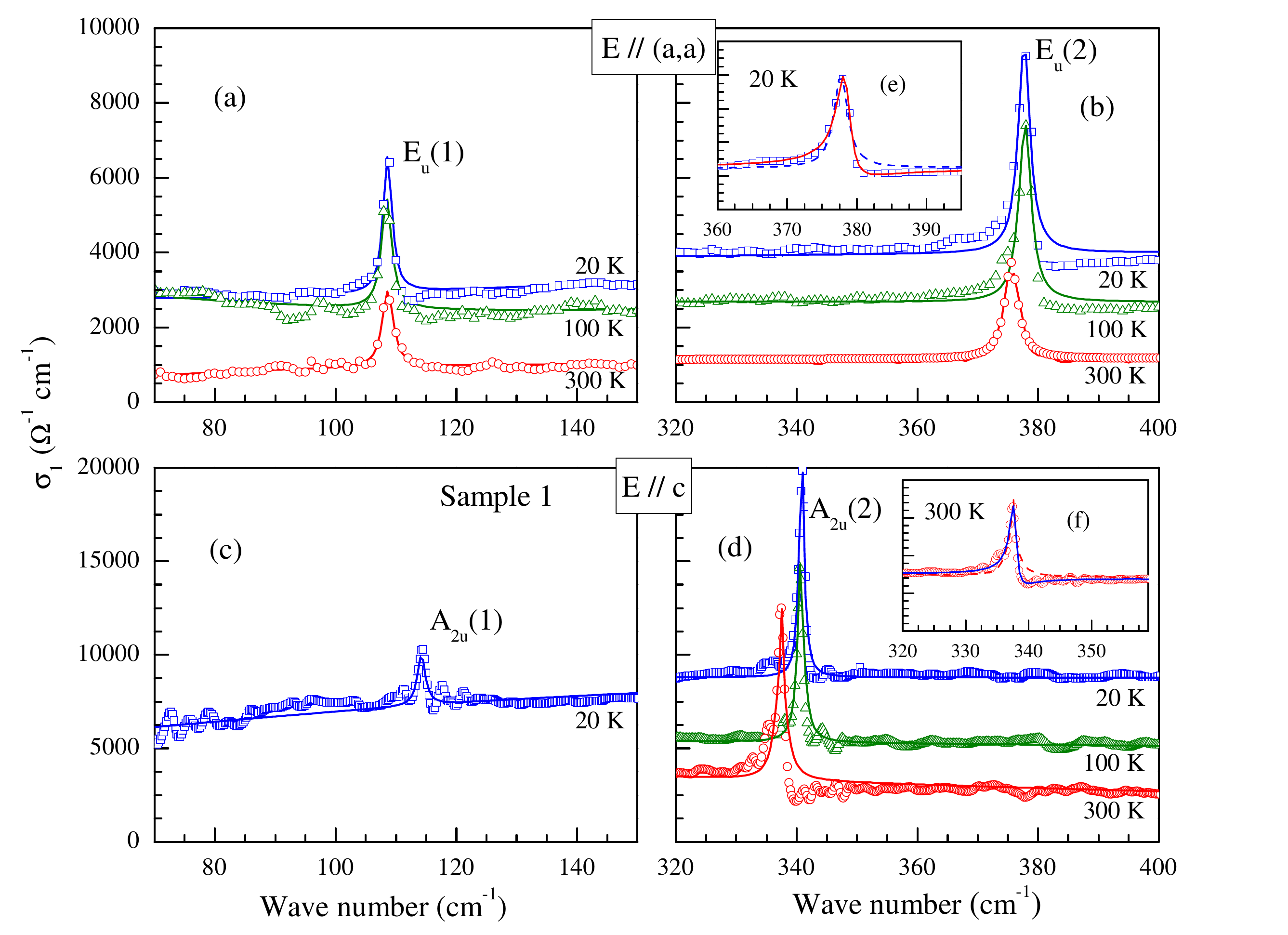}
\caption{(Color online) Optical conductivity centered around each phonon
of URu$_2$Si$_2$ at selected temperatures. Panels (a) and (b) are for light
polarized on the \paa\ plane and panels (c) and (d) are for measurements
along the \pc-direction. Symbols are the data and the lines are Drude-Lorentz
fits as described in the text. Inset (e) shows the highest frequency E$_u$(2)
mode at 20 K described by either a Lorentz oscillator (dashed line) or
a Fano mode (solid line). Panel (f) is the equivalent to panel (e) for
the highest frequency A$_{2u}$(2) mode at 300 K.}
\label{Sig1}
\end{figure*}

Figure \ref{Sig1} shows the optical conductivity around each phonon
for both polarizations. We do observe all predicted modes by group theory:
2 E$_u$ phonons in the \paa\ plane [panels (a) and (b)] and
2 A$_{2u}$ modes along the \pc-direction [panels (c) and (d)].
This is the first observation of the very weak A$_{2u}$ phonon at 115\icm, which
we could only detect at low temperatures. Above $\sim 30$~K, this phonon
becomes too broad to be resolved in the spectra.

In panels (a) through (d) the symbols are the data and the solid lines
are fits utilizing a Drude-Lorentz approach (Eqs. \ref{Drude} and \ref{Lorentz}). To describe the continuum,
we fitted the data to one Drude peak and two broad Lorentz oscillators.
This is a convenient way to parametrize the continuum but the values of
the parameters do not carry a particular physical meaning and will not
be discussed here. On top of this continuum we added a Lorentz oscillator
for each phonon. One can see that the Lorentz oscillator describes
reasonably well the phonon responses although the line shape is not
perfect at a few temperatures, such as 20 K in panel (b) and 300~K in
panel (d) (Fano line shape will be discussed later). Nevertheless, the Lorentz oscillator is very useful in analyzing
the phonon spectral weight.

The spectral weight, characterizing the charge in a restricted spectral
range, is defined as:

\begin{equation}
S_a^b = \int_{\omega_a}^{\omega_b} \sigma_1(\omega) \, d\omega \, .
\label{SW}
\end{equation}
When $\omega_a \rightarrow 0$ and $\omega_b \rightarrow \infty$, one
recovers the $f$-sum rule $S = (\pi / 2) (n e^2 / m)$ ($n$ is the
total number of electrons, $e$ is the electronic charge,
and $m$ the bare electronic mass). This rule states that the total
integral under the real part of the optical conductivity is a constant
independent of external parameters such as the temperature or pressure.

The spectral weight for a phonon within the Lorentz framework is:
\begin{equation}
S_p = \frac{\pi^2}{Z_0} \Delta \varepsilon \Omega^2 \, .
\label{SWPhonon}
\end{equation}
If phonons were decoupled from each other and from other excitations (\textit{e.g.} electronic continuum), Eq.~\ref{SWPhonon} should be temperature independent
for each phonon.

\begin{figure}
\includegraphics[width=0.9\columnwidth]{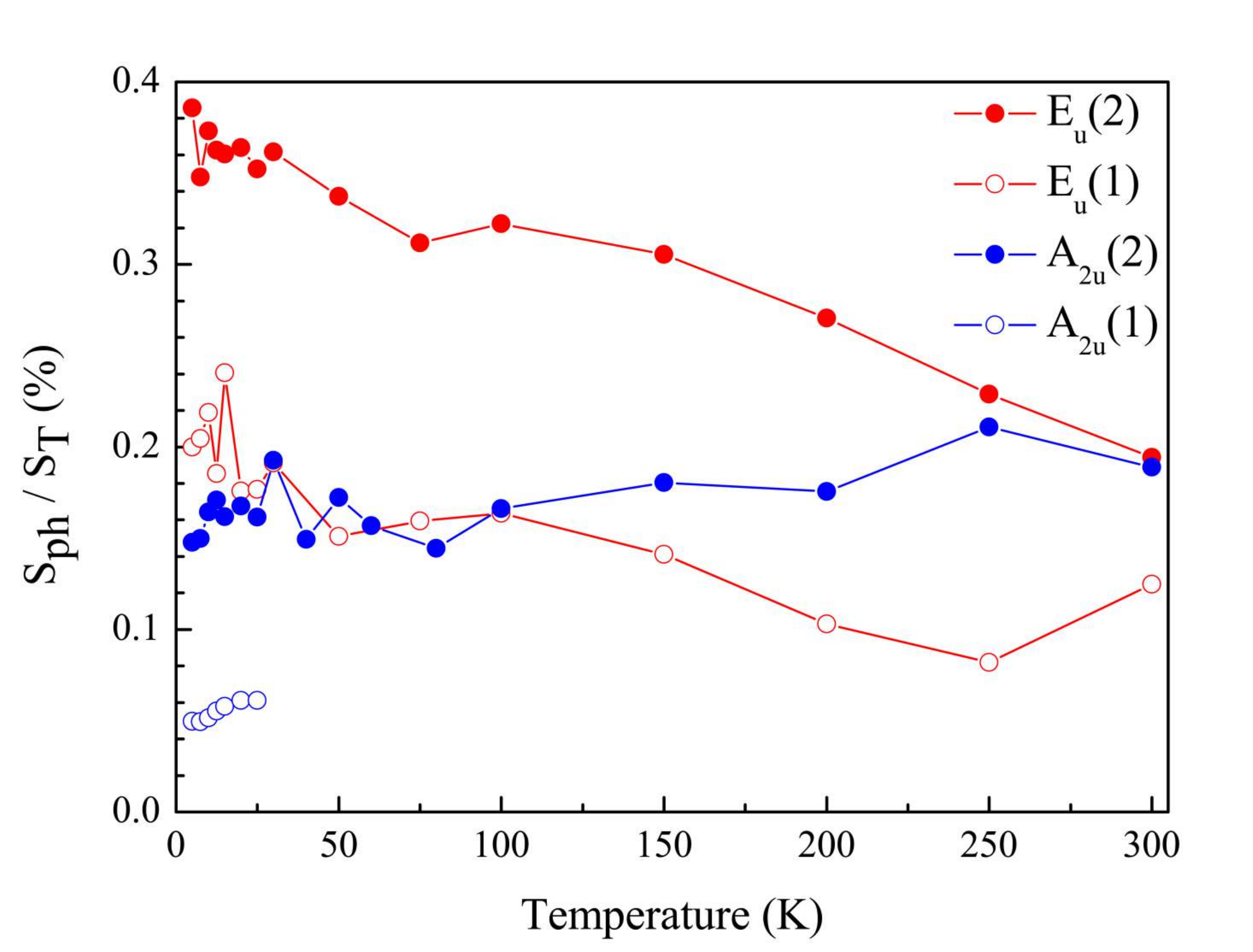}
\caption{(color online) Spectral weight of each phonon normalized by the
spectral weight up to $2\,000\icm$ calculated at room temperature.}
\label{SumRule}
\end{figure}

Figure \ref{SumRule} shows the temperature dependence of the spectral
weight of each phonon from Eq.~\ref{SWPhonon}, normalized by the total spectral weight for its
respective polarization at room temperature, integrated up to $2\,000\icm$.
Data for the lowest frequency A$_{2u}$ is shown for completeness, but we do not
have enough temperatures to draw any conclusion about this phonon.

The spectral weight of both E$_u$ phonons almost doubles upon
cooling the sample from room temperature to 5 K.
The highest energy A$_{2u}$ phonon also shows
a temperature dependent spectral weight, albeit of smaller magnitude.
Interestingly, for this phonon, the spectral weight decreases with
decreasing temperature.

A temperature dependent spectral weight indicates
that the effective charge of the phonons change with temperature.
As the sum rule states that the total spectral
weight must be conserved, this charge must be transfered from or to some
other excitation. Because the spectral weight of the phonons correspond
to less than 1\% of the total spectral weight, we do not have enough
resolution to pinpoint the energy (and hence the excitation) from which
this charge is being transfered. However, the obvious candidate is the
electronic continuum.

Indeed, let us go back to Fig.~\ref{Sig1} and make a closer inspection
on the phonon line shapes. In panel (e), we show the data for the highest
E$_{2u}$ phonon fitted by a Lorentz oscillator (dashed line) or a Fano mode
(solid line) at 20 K. Only the latter properly describes the asymmetry observed in
the measured data. This is the hallmark of an electron-phonon coupling and
substantiates our claim that the phonon charge is changing due to a
spectral weight transfer with the electronic continuum. Panel (f) shows
that the same effect is present in the highest energy A$_{2u}$ mode at
300 K.

\begin{figure}
\includegraphics[width=1\columnwidth]{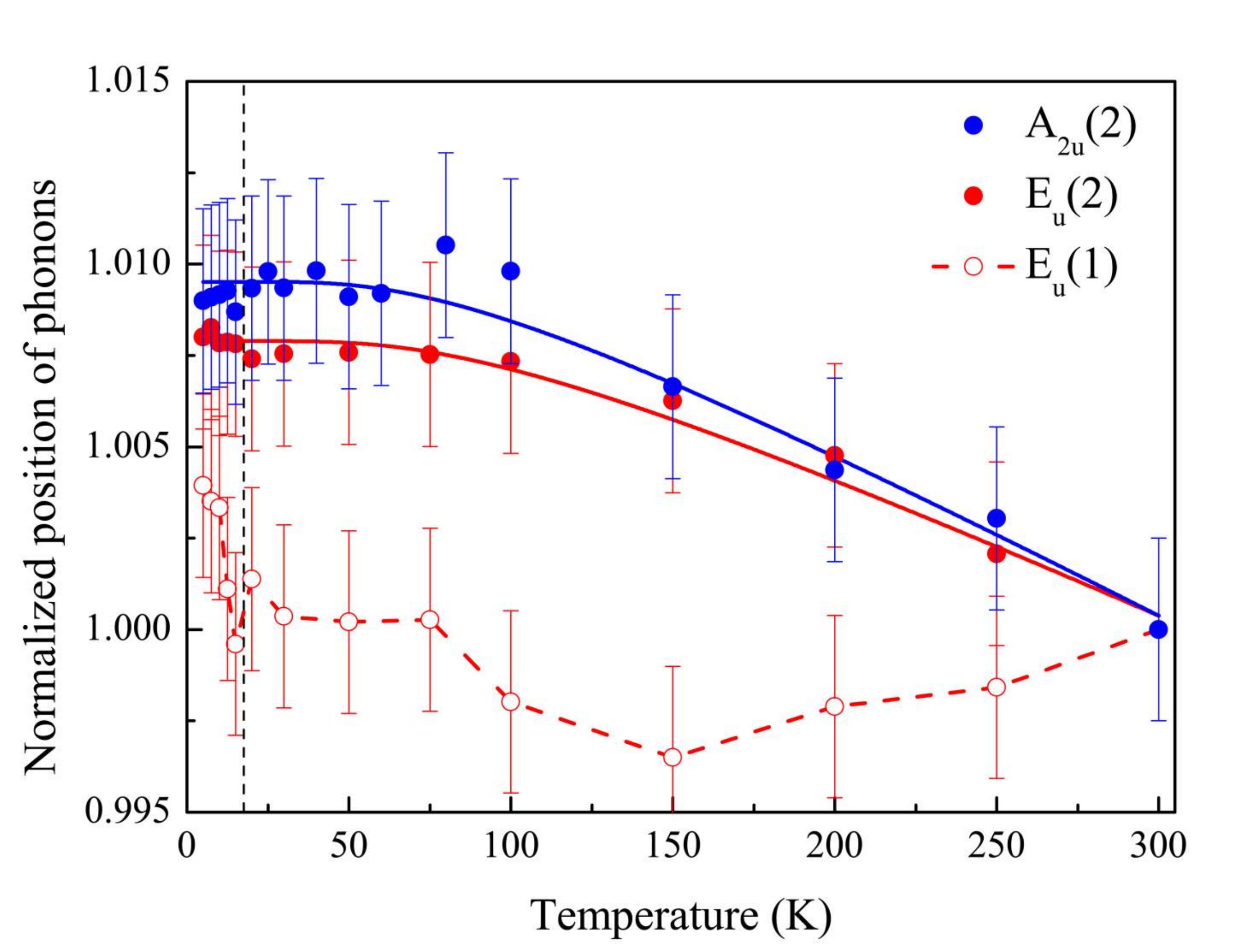}
\caption{(Color online) Temperature dependence of the frequency of A$_{2u}$(2), E$_u$(1) and E$_u$(2) phonon modes normalized to their value at 300~K. Lines are fits with a single anharmonic model (see Section \ref{raman:results}). The vertical black dashed line marks the transition into the hidden order.}
\label{TdepIR}
\end{figure}

Using the Fano mode fitting, we first extract the temperature dependence of the phonons energy normalized to their value at 300~K as presented Figure \ref{TdepIR}. At 300~K, the two E$_u$ phonon modes are seen at 108.7 and 375.6 cm$^{-1}$, and the highest A$_{2u}$ phonon mode at 337.8 cm$^{-1}$. Whereas the E$_u$(2) and A$_{2u}$(2) phonon modes exhibit the expected hardening when cooling down, the E$_{u}$(1) phonon mode shows constant energy down to $\sim$20K and a small hardening of $\sim$0.2~\% upon entering the HO phase. Even if only this low energy phonon E$_u$(1) has a singular temperature dependence, the E$_u$(2) and A$_{2u}$(2) phonon modes exhibit Fano line shape with peculiar temperature dependence, again evidencing that a complex electron-phonon coupling is in play in URu$_2$Si$_2$.

\begin{figure}
\includegraphics[width=0.8\columnwidth]{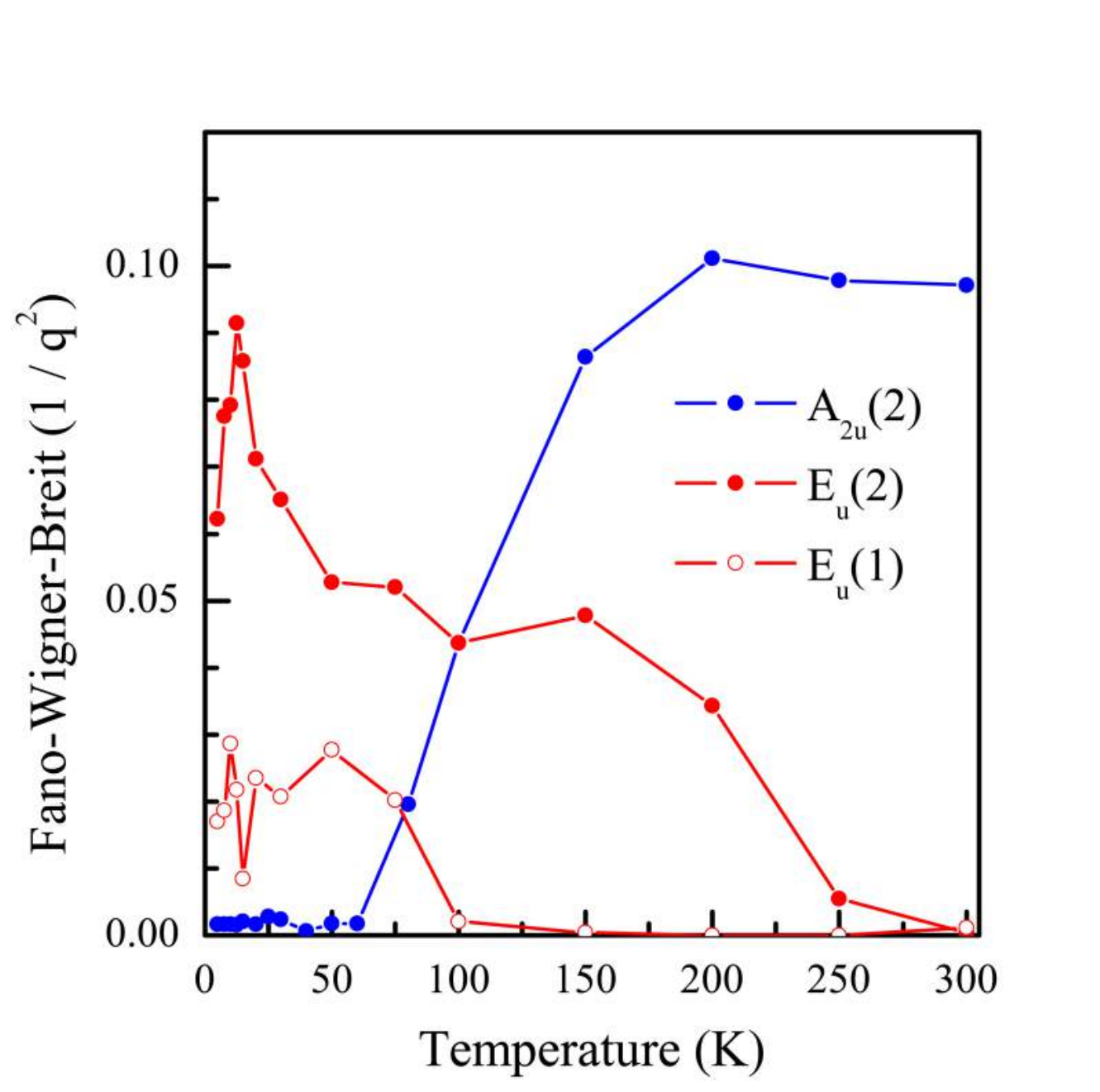}
\caption{(color online) Temperature dependence of Fano-Wigner-Breit
$q^{-2}$ parameter for both E$_u$ and the highest frequency A$_{2u}$ phonons. When an electron-phonon coupling occurs, the Fano-Wigner-Breit $q^{-2}$ parameter becomes different from zero.}
\label{FigFano}
\end{figure}

Fig.~\ref{FigFano} shows the temperature behavior for the Fano-Wigner-Breit
parameter for both E$_u$ ((1) at 108\icm\ and (2) at 378\icm) and the
highest energy A$_{2u}(2)$ (at 340\icm) phonons. The $q^{-2}$
parameter for mode E$_u(2)$ shows a behavior similar to its effective charge.
Both quantities increase almost featurelessly with decreasing temperature. This joint
behavior corroborates the electron-phonon coupling for this mode. Whereas we observe a clear drop in $q^{-2}$ at the hidden order transition for the mode E$_u$, a drop which is directly related to the loss of carriers number at T$_0$, we cannot pinpoint a particular change of coupling in the paramagnetic state.

The phonons E$_u(1)$ and A$_{2u}(2)$, on the other hand, show a striking change
of regime close to the Kondo transition where coherent transport develops. E$_u(1)$ has a
very small, yet finite, $q^{-2}$ at high temperatures. Close to the Kondo temperature
it suddenly increases and the phonon develops an asymmetric shape. The opposite is
observed for phonon A$_{2u}(2)$, which has an asymmetric shape and a large $q^{-2}$ value
above the Kondo temperature.

The temperature dependence of $q^{-2}$ probes the variation of the
continuum density of states close to the phonon energy. A decrease in
$q^{-2}$ indicates then that the density of states close to the phonon
energy becomes smaller, \textit{i.e.}, gapped. The optical conductivity results on the A$_{2u}$ mode indicate that we would have a gapped system along $k_z$ around 40 meV in the Kondo regime. Phonon E$_u(1)$, conversely, may indicate that the system is gapped or with strongly depleted density of states on the $(k_x, k_y)$ plane at about 12 meV but this gap closes below the Kondo temperature or more generally the electronic density of state is enhanced when entering into the Kondo liquid regime.

Such observation of a temperature dependent Fano shape of a phonon mode in metallic Kondo systems has already been reported in CeCoIn$_5$ by Raman scattering\cite{martinho_vibrational_2007}. Indeed, according to their study of the lattice dynamics and electronic Raman response of CeCoIn$_5$, the entrance into the Kondo liquid regime below the crossover temperature T*$\sim$~45~K manifests by the divergence of the Fano coefficient of the A$_{1g}$ phonon and by the significant drop of the scattering rate of the electronic scattering background. Both have been related to the enhancement of the electronic density of states due to the hybridization of 4f electrons with the conduction band. However the behavior as measured on the A$_{2u}(2)$ mode in URu$_2$Si$_2$ and even more the concomitant opposite behaviors of two phonons A$_{2u}(2)$ and E$_u(1)$ within the same compounds is striking and has never been reported to our knowledge. Most probably, this points to strongly momentum-dependent Kondo physics in URu$_2$Si$_2$, which affect distinctively both phonon modes with movements into different and perpendicular planes, namely the (x,y) and (z) planes.

\section{Phonons and magnetic excitations studies by inelastic neutron scattering}
\label{INS}

\subsection{Experimental Details}

The phonon spectrum of URu$_{2}$Si$_{2}$ was investigated by Inelastic Neutron Scattering (INS) at the Institute Laue-Langevin.
The first experiment was performed on the thermal neutron three axis spectrometer IN8. In the first configuration, the initial beam is provided by a double focusing Si monochromator (Si(1,1,1)) and the scattered beam is analyzed by a double focusing Pyrolytic Graphite (PG) analyzer (PG(0,0,2)) with fixed $k_{F}$= 2.662 \AA$^{-1}$. In the second configuration, the initial beam is provided by a double focusing Cu monochromator (Cu(2,0,0)) and the scattered beam is analyzed as previously but with fixed $k_{F}$= 4.1 \AA$^{-1}$. This second configuration is used to investigate high energy modes.  The second experiment was performed on the thermal neutron three axis spectrometer IN22 where the initial beam is provided by a double focusing PG monochromator (PG(0,0,2)) and the scattered beam is analyzed by a double focusing PG analyzer (PG(0,0,2)) with fixed $k_{F}$= 2.662 \AA$^{-1}$. For these two experiments, the sample is a cylinder of diameter 4.5 mm and of length 8 mm along the $a$-axis; the scattering plane is defined by (1,0,0) and (0,0,1). The third experiment was performed on IN22 in polarized neutron setup with Heussler monochromator and analyzer with fixed $k_{F}$= 2.662 \AA$^{-1}$. The neutron polarization is kept along the neutron path by guide fields and by an Helmholtz coil around the sample ; a Mezei flipper is placed before the analyzer. The experiment was performed with the neutron polarization parallel to the scattering vector ($\bf{P}$ // $\bf{Q}$). With this configuration, all the magnetic scattering appears in the Spin Flip (SF) channel while the phonon scattering appears in the Non Spin Flip (NSF) channel. For this experiment, the sample is a cylinder of diameter 4.5 mm and of length 12 mm along the $c$-axis and the scattering plane is defined by (1,0,0) and (0,1,0).
In all measurements, the sample was inside an helium-4 flow cryostat covering the range 2-300 K. A PG filter was always placed in the scattered beam in order to remove higher order contamination. (Note for comparison between Raman, IR and neutron scattering that 1~meV=8.06~cm$^{-1}$).

\subsection{RESULTS}

\begin{figure}
\centering
\includegraphics[width=1\linewidth]{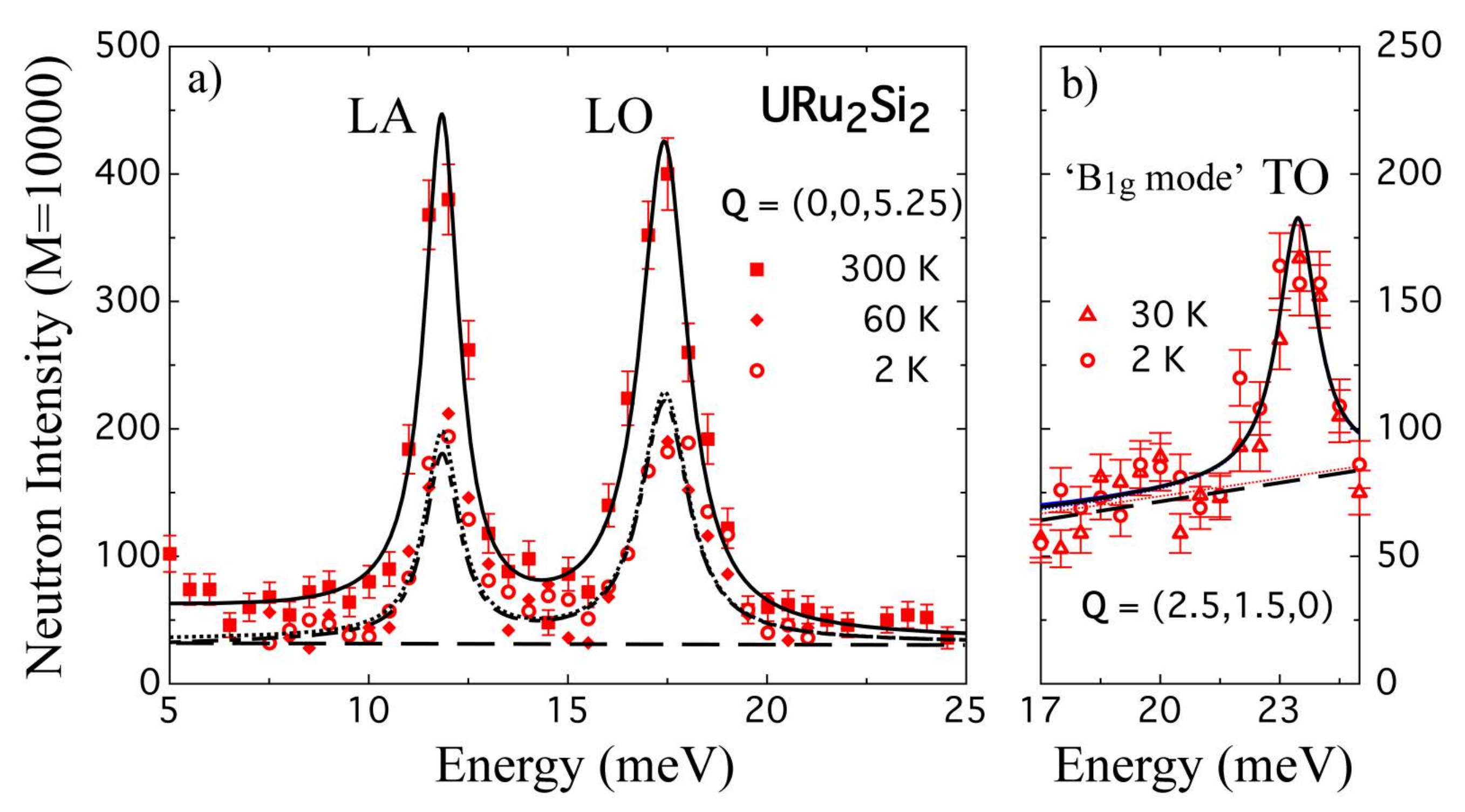}
\vspace{-0.5cm}
\caption{(Color online) Inelastic Neutron Scattering phonon spectra measured for (a): $\bf{Q}$=(0, 0, 5.25) at $T$ = 2, 60 and 300 K performed on IN8 and (b): $\bf{Q}$=(2.5, 1.5, 0) at $T$ = 2 and 30 K performed on IN22. L and T, A and O stand for longitudinal and transverse, acoustic and optic, respectively. The TO mode belongs to the branch that corresponds to the B$_{1g}$ mode at $\bf{Q}$=0. Lines are fits of the data with an harmonic oscillator lineshape and sloppy background. The neutron intensity is given for a normalized incident flux (M) which corresponds to a counting time of approximatly 80s.}
\label{fig6}
\end{figure}

\begin{figure}
\centering
\includegraphics[width=1\linewidth]{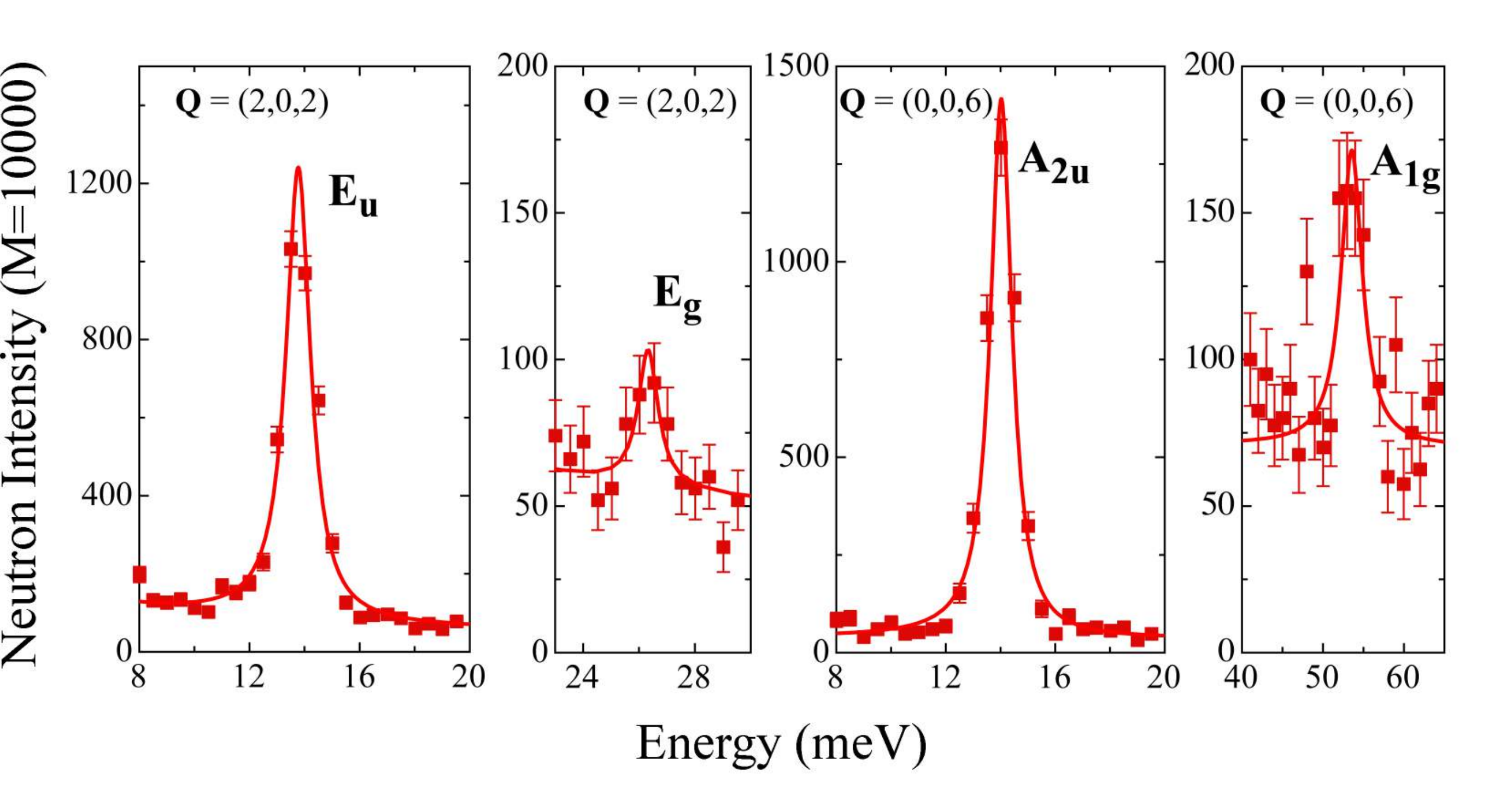}
\vspace{-0.5cm}
\caption{(Color online) Inelastic Neutron Scattering phonon spectra at $\Gamma$ point measured for $\bf{Q}$=(2, 0, 2) and $\bf{Q}$=(0, 0, 6) at $T$ = 300K. The E$_{u}$ and E$_{g}$ phonon modes have been observed for $\bf{Q}$=(2, 0, 2) and A$_{2u}$ and A$_{1g}$ phonon modes for $\bf{Q}$=(0, 0, 6). The neutron intensity is given for a normalized incident flux (M) which corresponds to a counting time of approximately 80s for E$_{u}$, E$_{g}$, A$_{2u}$ and 110s for A$_{1g}$. Lines are fits of the data with an harmonic oscillator lineshape.}
\label{fig7}
\end{figure}

\begin{figure}
\centering
\includegraphics[width=1\linewidth]{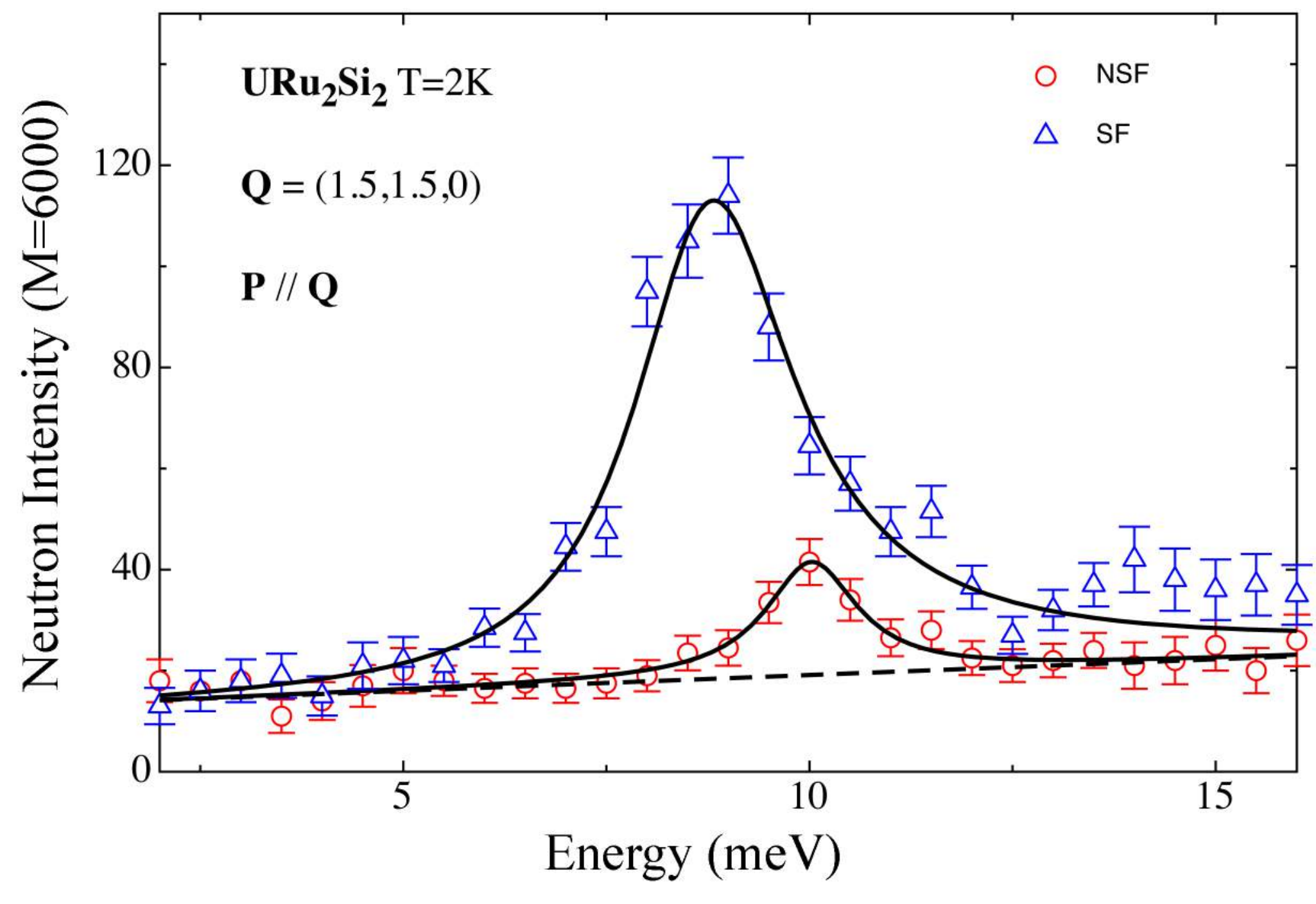}
\vspace{-0.5cm}
\caption{(Color online) Polarized Inelastic Neutron Scattering spectra measured for $\bf{Q}$=(1.5, 1.5, 0) in the Non Spin Flip (NSF) and Spin Flip (SF) channels for $T$ =  2 K. Full lines are fits of the data with damped harmonic oscillator lineshape for the phonon and $\omega$-lorentzian lineshape for the magnetic excitation (For such a lineshape, see e.g.~\cite{panarin_effects_2011}). The dashed line indicates the background. The neutron intensity is given for a normaized incident flux which corresponds to a counting time (M) of approximatly 25min.}
\label{fig8}
\end{figure}

In the present paper, the scattering vector $\bf{Q}$ is decomposed into $\bf{Q}=\bf{G}+\bf{q}$, where ${\bf{G}}$ is a reciprocal lattice wave vector and $\bf{q}$ is the wave-vector of the excitation. The cartesian coordinates of $\bf{q}$=($h$, $k$, $l$) are expressed in reciprocal lattice units (r.l.u.). Representative phonon spectra measured on IN8 are shown in Figure~\ref{fig6}(a) for $\bf{Q}$=(0, 0, 5.25). The two peaks at 11.8 and respectively 17.5 meV correspond to longitudinal acoustic (LA) and respectively longitudinal optic (LO) modes. Fits of the data are made at $T$=300 K using a flat background and a damped harmonic oscillator for the scattering function. The data at 2~K and 60~K are described by using the parameters obtained at 300~K except the intensity that is rescaled by the Bose factor. This procedure allows us to spot anomalous temperature behavior of the phonons.
\par
The overall phonon modes measured along [0,0,1], [1,0,0] and [1,1,0] directions do not show noticeable temperature dependence on cooling from 300 K to 2 K or on crossing $T_{0}$, except for a small expected hardening (See the LO branch in Figure~\ref{fig6} shifting a little to higher energy with decreasing temperature) which is a normal behavior of phonon on cooling.
Although the softening of the B$_{1g }$ mode seen by Raman scattering is too small (0.5\%) to be detected by neutron measurement, a larger softening could occur at finite $\bf{q}$. Therefore a particular attention has been focused on the temperature dependence of the "B$_{1g}$" branch for $\bf{q}\neq0$ in [0,0,1] and [1,1,0] directions. As presented for instance Figure~\ref{fig6}(b) for $\vec Q$=(2.5,1.5,0), no temperature difference has been observed between 2K and 30K.
\par
Figure~\ref{fig7} shows the four phonon modes E$_{u}$, E$_{g}$, A$_{2u}$ and A$_{1g}$ seen at $\Gamma$ point by INS. A good agreement is found with the energy of these phonon modes measured by IR and Raman spectroscopy.

\par
Further emphasis was given in the study of the phonons along the [1,1,0] direction by polarized neutron scattering following the report of anomalous phonon softening in this direction below $T_{0}$ \cite{butch_soft_2012}.

\begin{figure*}[htbp]
\centering
\includegraphics[width=0.9\linewidth]{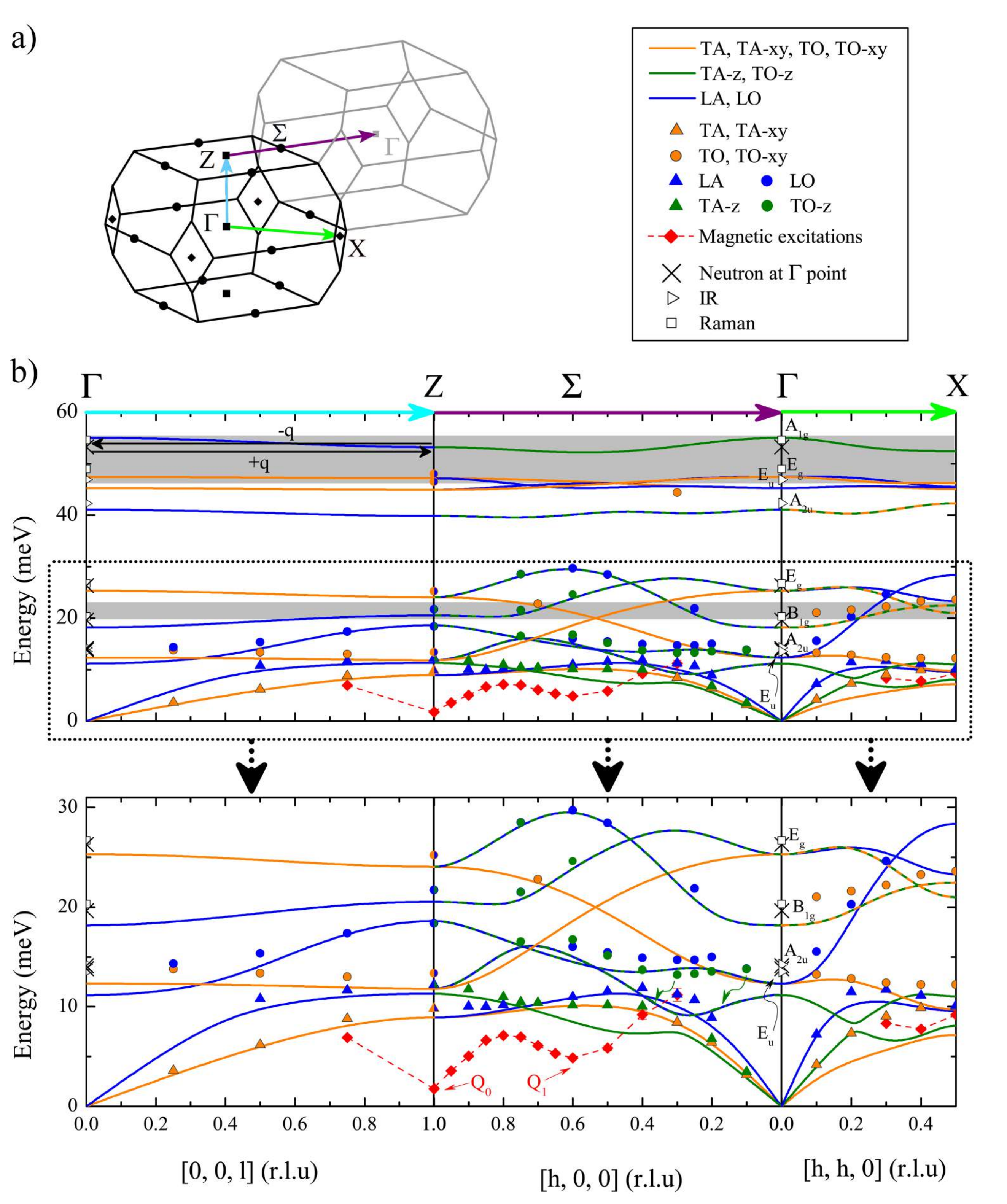}
\caption{(Color online) (a) Brillouin zone of the body centered tetragonal crystal structure of URu$_2$Si$_2$ in the paramagnetic state. The arrows indicate the direction of the dispersion for the Inelastic Neutron Scattering (INS) measurements, as reported below. (b) Phonons and magnetic excitations (full red diamonds) dispersion along the $\Gamma$Z ([0,0,1]), Z$\Sigma\Gamma$ ([1,0,0]) and $\Gamma$X ([1,1,0]) directions. Full circles and full upward triangles are INS data for the optical and acoustic branches, respectively. Empty symbols are data obtained in the zone center at $\Gamma$, with cross, open triangles and squares for INS, IR and Raman measurements, respectively. Dashed lines are guides to the eyes for the magnetic excitations dispersion. Solid lines corresponds to the $\it{ab~initio}$ GGA calculation with spin-orbit coupling and u=0.06 \AA~ (See section \ref{sec:theory}). When the dispersion curve has a mixed character (for instance from LO (in $\Gamma$) to TO-z (in $Z$) along [1,0,0]), the solid lines are bicolored. The black arrow points to the E$_{u}(1)$ phonon mode measured by IR at $\Gamma$ point. The green arrows indicate the calculated branch which corresponds to the measured points (green full circles). The grey areas indicate the energy range for all possible double phonon processes observed by Raman scattering (see Section \ref{raman:results}). The two ($\pm\vec{q}$) black arrows for the [0,0,l] direction show one possible double phonon process on the "A$_{1g}$" branches. Q$_0$ and Q$_1$ correspond to the minima in magnetic excitation dispersion, known as commensurate and incommensurate excitations, respectively.}
\label{fig9}
\end{figure*}

Figure~\ref{fig8} shows a representative measurement performed on IN22, with the neutron polarization parallel to $\bf{Q}$, at X point in the Brillouin zone (See Figure~\ref{fig9}) for $\bf{Q}$=(1.5, 1.5, 0) at $T$=2 K for SF and NSF scattering. In the NSF scattering a phonon mode is observed at around 10 meV. In the SF channel, the large intensity peak centered at around 8.7 meV corresponds to the well-known magnetic excitation of URu$_{2}$Si$_{2}$. The peak position is in agreement with an early study performed along the [1,1,0] direction by Broholm et al. \cite{broholm_magnetic_1991}. We therefore conclude the reported soft phonon mode \cite{butch_soft_2012} is in fact a magnetic excitation, an unambiguous result obtained by polarized INS.

\par
Figure~\ref{fig9} summarize all the phonon and magnetic excitations dispersions we have measured along the main directions [1,0,0], [0,0,1] and [1,1,0]. As we have not observed any significant variation of the phonon energies as a function of temperature between 300~K and 2~K, we have reported the energy average of all measured temperature for each phonon mode.  There is a fairly good agreement between the measurement and the calculation which includes the spin-orbit coupling (See section~\ref{sec:theory}), except for few branches. The energy of the transverse (T) modes TA-z and TO-z (or "A$_{2u}$(1)" branch) along [1,0,0] and TA-xy along [1,1,0] lines is $\sim~20\%$ higher than what is expected by the calculation.
 \par
 We have not observed any magnetic excitations along [1,0,0] nearby $\Gamma$ point. Then, most probably, the excitations reported by Broholm \cite{broholm_magnetic_1991} in this $\vec{k}$-space zone are the optical phonon mode ('E$_u$' or 'A$_{2u}$' branches).
\par
Furthermore, we report no particular anomaly of the phonon branches around $Q_0$ and $Q_1$ points where magnetic excitations are centered. No strong magneto-elastic coupling related to these magnetic modes is in play in URu$_2$Si$_2$.\newline

\section{THEORETICAL CALCULATIONS OF PHONON DISPERSION CURVES}
\label{sec:theory}

The calculations have been performed using the density functional theory (DFT)
implemented in the VASP software. \cite{kresse_efficient_1996}
The electron potentials and wave-functions were obtained within the projector-augmented waves method \cite{blochl_projector_1994}
and the exchange and correlation energy was described by the generalized-gradient approximation. \cite{perdew_generalized_1996}
The expansion of the single-particle plane waves has been restricted by the energy cutoff of 340 eV.
The electronic and crystal structure have been optimized in the $2\times2\times1$
supercell (40 atoms) with the periodic boundary conditions.
We performed two types of calculations, with and without spin-orbit
coupling (SOC), assuming in both cases the non-magnetic ground state.
The optimization with the SOC gives slightly larger lattice constants
than without SOC (see Table~\ref{tab-lattice}). In both cases, the optimized lattice constants
and the position of Si atoms ($z$) show good agreement with the experimental values and the previous relativistic full-potential calculations \cite{oppeneer_electronic_2010}.

\begin{table}[h]
\caption{Lattice parameters obtained in calculations with and without SOC compared
with the experimental values at $T=4.2$ K taken from Refs. \cite{palstra_superconducting_1985} and \cite{cordier_structural_1985}.}
\renewcommand{\arraystretch}{1.5}
\begin{tabular}{|c|c|c|c|c|c|}
\hline
       & $a$ (\AA) & $c$ (\AA) &  $c/a$  & $V$(\AA$^3$) &  $z$ \\
\hline
SOC    &   4.143   &   9.589   &  2.31   &  164.59       &  0.374 \\
no SOC &   4.136   &   9.549   &  2.31   &  163.35       &  0.375 \\
exp.   &   4.124   &   9.582   &  2.32   &  162.96      &  0.371 \\
\hline
\end{tabular}
\label{tab-lattice}
\end{table}

\begin{table*}[!t]
\caption{Phonon energies, $E$, for the phonon modes of URu$_2$Si$_2$ in \icm at the $\Gamma$ point obtained by Raman, infrared and inelastic neutron scattering measurements at the lowest temperature and, by {\it ab initio} studies at 0~K. We report the energies calculated with the spin-orbit coupling and $u$=0.06 \AA. The ones calculated without the spin-orbit coupling and with $u$=0.06~\AA~ are presented in parentheses. On the right: calculated atomic intensities of URu$_2$Si$_2$ with spin-orbit coupling and displacements of $u=0.06$~\AA\ for the lattice modes at the $\Gamma$ point. The $\sum_{\mu,i} |\mathbf{e}_i(\mathbf{k},j;\mu)|^2$ filter has been used to calculate how much the atom $\mu$ is involved in the $j$th vibration. $i=x,y,z$ and, here  $k$=0.}
\begin{ruledtabular}
\begin{tabular}{cccccccc}
 Phonon modes           & \multicolumn{3}{c}{$E$ measured by}         &  $E$          & \multicolumn{3}{c}{Atomic intensities} \\
at $\Gamma$ point & Raman spectroscopy & IR &  neutron  &  calculated   &  U   & Ru   & Si   \\
  \cline{1-1} \cline{2-5} \cline {6-8}
A$_{2u}$ &  /  &  114.8    &  111.2   &   90.1 (84.7)   &   0.52     & 0.42  & 0.06 \\
E$_u$    &  /  & 109.1     &  113.6   &   99.5 (101.6)  &   0.52     & 0.42  & 0.06 \\
B$_{1g}$ & 163.6 &  /  & 158.1 (leakage) &   146.7 (144.2)& 0           & 1.00  & 0.00 \\
E$_g$    & 215.1 & /   & 212.0           &   204.3 (206.6)&  0          & 0.74  & 0.26 \\
A$_{2u}$ &  /     & 340.8     & /     &  331.5 (331.3)   &         0     & 0.17  & 0.83 \\
E$_u$    & /      & 378.6      & /    &   365.3 (364.2)  &          0      & 0.17  & 0.83 \\
E$_g$    & 394.1    & /      & /      &   382.9(382.7)   &         0      & 0.26  & 0.74 \\
A$_{1g}$ & 439.7       & /   & 429.1  &   443.9 (459.7) &         0     & 0.00  & 1.00 \\
\end{tabular}
\end{ruledtabular}
\label{tab2}
\end{table*}

The phonon dispersion curves were obtained by using the direct method.\cite{parlinski_first-principles_1997, parlinski_phonons_2011}
In this approach, the force constants are derived from the Hellmann-Feynman (HF) forces calculated $ab initio$ by displacing atoms from equilibrium positions.
Due to symmetry constraints, only three atoms (Si, Ru, and U) have to be
displaced along two non-equivalent directions, $x$ and $z$.
In total, 12 independent calculations have been performed,
including displacements in positive and negative directions.
The phonon dispersions were calculated by the exact diagonalization of the dynamical matrix,
obtained directly from the force constants.

The nature of the Raman and infrared-active vibrations are described in Fig.~\ref{fig1} and Table~\ref{tab1}. However, $ab initio$ calculations of the phonon dispersion curves allow for a more detailed description of the actual atomic character of the vibrations. The atomic intensities for each phonon branch at the $\Gamma$ point is shown in Table~\ref{tab2}. In this case, the intensity refers to the square of the vibrational amplitude of each atom for a given branch. This is
particularly useful for the infrared-active A$_{2u}$ and E$_u$ modes, which in principle involve
displacements of all of the atoms in the unit cell. The atomic intensity reveals that at the $\Gamma$ point, the low-frequency A$_{2u}$ and E$_u$ modes involve mainly the U and Ru atoms, while the high-frequency A$_{2u}$ and E$_u$ modes are almost entirely Si in character with a slight involvement of the Ru atom.

\begin{figure}[!h]
\centering
\includegraphics[width=1\linewidth]{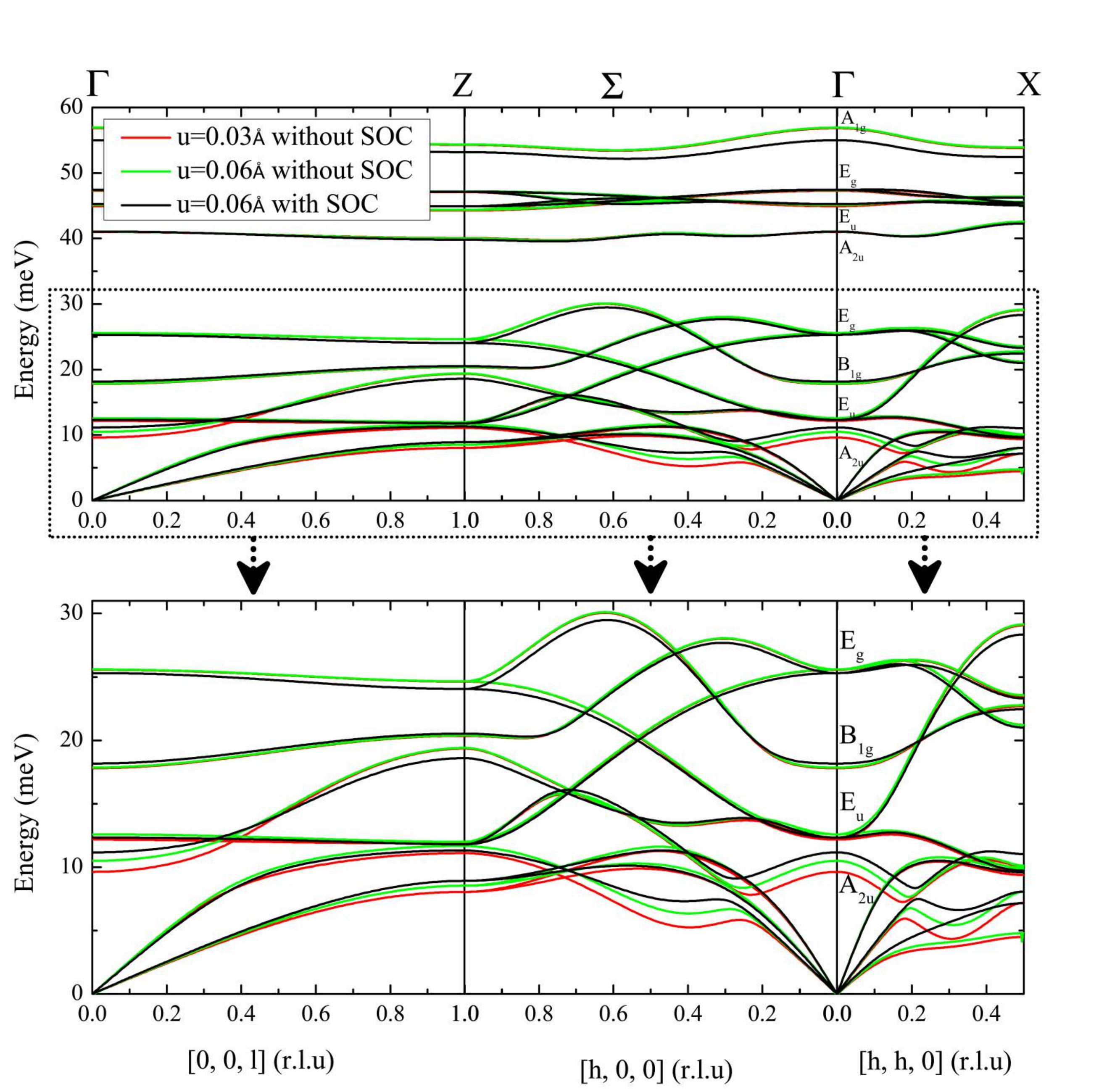}
\vspace{0cm}
\caption{(Color online) {\it Ab initio} calculation of the phonon dispersion curves along $\Gamma Z$, $\Gamma \Sigma Z$ and $\Gamma X$ at 0~K. The parameter $u$ sets for the displacements of the atoms. Increasing $u$ from 0.03~\AA~ (red line) to 0.06~\AA~ (green line) is used to probe the effects of anharmonicity. Black line shows the calculation with spin-orbit coupling (SOC). The black line is the closest to the experimental results and is the one reported in Figure\ref{fig9}}.
\label{fig10}
\end{figure}

In order to investigate possible anharmonic effects, two different sets of displacements,
with $u=0.03$ \AA~ and $u=0.06$ \AA, were used to derive the HF forces. This approach would give us information about the possible
deviation from the harmonic potential and it has been used previously to study the anharmonic
behaviour in magnetite \cite{hoesch_anharmonicity_2013}. The results of both calculations are presented in Figure~\ref{fig10}.
Instead of phonon softening typical for an anharmonic potential, we observe the increase in energy
at larger value of $u$. The effect is more apparent for the TA-z and the lowest TO mode along
the $\Gamma X$ and $\Gamma\Sigma Z$ directions and TA-xy along the $\Gamma Z$ direction.
At the $\Gamma$ point, two lowest infrared modes A$_{2u}$ and E$_u$ shift by $+7\%$ and $+3\%$,
respectively~\footnote{In the case of smaller displacements $u=0.03$ \AA\ numerical values of forces
are too small, which may be the origin of some numerical error, therefore, we present phonon energies c
and dispersions obtained for $u=0.06$ \AA.}. The energies of other modes depend
on $u$ very weakly.

To analyze the effect of the SOC, we have compared the results obtained with and without
the SOC calculated for $u=0.06$ \AA. As we see in Figure~\ref{fig10}, the strongest effect
is found for the lowest infrared A$_{2u}$ mode, which is shifted upward by $6.5 \%$ due to the SOC.
The increase in energy in spite of larger lattice constants indicates a direct influence of the modified electronic structure on interatomic forces and phonon energies.
Interestingly, the modes, which are strongly modified by the SOC, exhibit also
the most pronounced dependence on $u$, and simultaneously they show the largest disagreement
with the INS data (see Figure~\ref{fig9} and Table~\ref{tab2}). The results obtained with the SOC and $u=0.06$ \AA,are slightly closer to the experimental points than the two other calculations.

\section{General discussion and Conclusion}

Recently, orthorhombic distortion upon entering the HO has been measured by X-ray scattering by Tonegawa et al. \cite{tonegawa_direct_2014} in disagreement with results by Amitsuka et al. \footnote{H. Amitsuka, \textit{Workshop on hidden order, superconductivity and magnetism in URu$_2$Si$_2$}, Leiden (2013).}. Quantitative prediction of the effect of such distortion on the lattice dynamics would be of high interest as the phonon modes with displacements in the (a,a) plane would be expected to broaden or split. We do not observe any splitting or broadening of the phonons measured by Raman scattering or by optical conductivity measurements, but a quantitative comparison with the predictions based on this recent measurement and the width across T$_0$ as measured by optical spectroscopy would be necessary to definitively conclude.

Based on the observation of the same characteristic vector Q$_0$ in the HO and AF phase \cite{hassinger_similarity_2010,yoshida_translational_2013,buhot_a2g_2014} (Brillouin zone folding from a bct to st) and the absence of lattice distortion across T$_0$, Harima et al. \cite{harima_why_2010} have selected 4 subgroups of the group 139 as candidates for the lower space group of the HO state (n$^\circ$ 126, 128, 134, and 136); all have D$_{4h}$ symmetry. Even without lattice distortion, across T$_0$, group theory predicts that new active phonon modes are allowed to emerge (B$_{2g}$ phonon mode in the group n$^\circ$ 126, 134 and 136; only the group n$^\circ$ 128 doesn't have active B$_{2g}$ phonon mode), some modes can be split (E$_g$ or E$_u$ modes) or new atoms are allowed to participate to the phonon movements (Uranium atoms in the E$_g$ mode in group n$^\circ$ 126 and 134). None of these predictions have been observed here. Of course the effects of the electronic transition on the lattice dynamics might be very limited. Quantitative calculations on the lattice dynamics based on precise electronic ordering at T$_0$ would be necessary to distinguish which effects would be sizable.

In conclusion, we have performed Raman scattering, optical conductivity, inelastic neutron scattering measurements and {\it ab initio} calculations focused on the lattice dynamic properties of URu$_2$Si$_2$ in the 300~K - 2~K temperature range. We have measured all the optical phonon modes at the center of the Brillouin zone (BZ) and we have followed almost all phonon branches below 30~meV in the main symmetry directions of the BZ, together with their temperature dependencies. No particular effect of the entrance into the hidden order state has been detected except a change in the Fano shape of the E$_{u}$(2) phonon mode, a phonon which exhibits also a large increase of its spectral weight upon cooling from 300~K consistently with important electron-phonon coupling for this phonon. We attribute this behavior to the large loss of carriers upon entering the HO state. Other main effects have been obtained when entering into the Kondo regime. Indeed, we measure a small (0.5$\%$) but sizable softening of the B$_{1g}$ phonon mode below $\sim$100~K. Most probably a complex electron-phonon coupling is in play, related to the Kondo physics. This and the previously reported softening of the elastic constant of the same symmetry observed by ultrasound velocity measurements strongly suggest a B$_{1g}$ symmetry-breaking instability in the Kondo regime. The Kondo cross-over also impacts the infrared-active E$_u$(1) and A$_{2u}$(2)modes. Both of them present a Fano shape but whereas the A$_{2u}$(2) mode loses its Fano shape below 150~K, the E$_u$(1) mode acquires it below 100~K, in the Kondo cross-over regime. We attribute this behavior to strongly momentum-dependent Kondo physics.
By drawing the full dispersion of the phonon modes and magnetic excitations, we conclude that there is no strong magneto-elastic coupling in URu$_2$Si$_2$. No remarkable temperature dependence has been obtained by INS including through the hidden order transition. Thanks to polarized inelastic neutron scattering, we were able to distinguish between phonon and magnon modes near the X and $\Gamma$ points of the BZ, shedding light on previous reports \cite{broholm_magnetic_1987, butch_soft_2012}. The {\it ab initio} calculations of phonon energies and polarization vectors allowed us for the detailed analysis of phonon modes in the zone center and along the high-symmetry directions. A good agreement between the theory and experiment observed for most of dispersion curves indicates the itinerant character of $5f$ electrons. The discrepancy found for the lowest TA and TO modes propagating in the $(a,b)$ plane may be caused by additional effects such as strong electron correlations, magnetic interactions or relativistic effects not fully included in the present calculations.

\begin{acknowledgments}

This work was supported by the Labex SEAM (Grant No. ANR-11-IDEX-0005-02) and by the french Agence Nationale de la Recherche (ANR PRINCESS). CCH is supported by the U.S. Department of Energy (DOE), Office of Basic Energy Sciences,  Division of Materials Sciences and Engineering under Contract No. DE-AC02-98CH10886. The IT4Innovations National Supercomputing Center, VSB-Technical University, Ostrava, Czech Republic is acknowledged for providing the computer facilities under Grant Reg. No. CZ.1.05/1.1.00/02.0070.
We thanks I. Paul, G. Knebel, C. Lacroix and P. Oppeneer for very fruitful discussions.

\end{acknowledgments}

\bibliographystyle{apsrev4-1}
\bibliography{biblio}

\end{document}